\documentclass[a4paper,11pt]{article}

\usepackage{jheppub}
\usepackage[english]{babel}
\usepackage[dvipsnames]{xcolor}
\usepackage[T1]{fontenc}
\usepackage{cite}
\usepackage{relsize,scalerel}
\usepackage[export]{adjustbox}
\usepackage{amsmath, amsfonts, amssymb,wasysym}
\usepackage{slashed}
\usepackage{simpler-wick}

\usepackage{graphicx,tikz,pgf}
\usetikzlibrary{cd,shapes,decorations.pathmorphing,decorations.markings,decorations.pathreplacing}
\usepackage{ifthen,xparse}

\definecolor{BlobLineColour}{RGB}{96, 96, 96} 
\definecolor{BlobFillColour}{RGB}{96, 96, 96}  
\definecolor{GluonColour}{RGB}{64, 64, 64} 
\definecolor{DoubleGluonColour}{RGB}{64, 64, 64} 
\definecolor{GhostColour}{RGB}{64, 64, 64} 
\definecolor{QuarkColour}{RGB}{64, 64, 64} 
\definecolor{PlainColour}{RGB}{64, 64, 64} 
\definecolor{DoublePlainColour}{RGB}{64, 64, 64} 
\definecolor{ChargedScalarColour}{RGB}{64, 64, 64} 
\definecolor{NeutralScalarColour}{RGB}{64, 64, 64}

\def\picSc{0.55} 
\def\blobSc{0.5} 
\def\bigBlobSc{0.75}

\def\lineW{0.5} 

\ExplSyntaxOn
\NewDocumentCommand{\newwhiledo}{m m}
  {
   \bool_while_do:nn { \int_compare_p:n {#1} } { #2 }
  }
\ExplSyntaxOff

\newcommand\blobNode[4]{\node[circle,line width=\lineW mm,BlobLineColour!80,draw,fill=BlobFillColour!10] at (#1,#2)[scale=#3] {#4};}

\newcommand\leftArc[2]{
	\pgfmathsetmacro{\np}{#1} 
	\ifthenelse{\equal{\np}{1}}	{
		\def\angle{180}
		\node[circle,line width=\lineW mm,BlobLineColour!80,draw,fill=BlobFilleColour!10] (V\np) at (\angle: 1)[scale=\blobSc] {#2};
	}	{
		\foreach \k in {1,...,\np}
			\def\angle{{ 360 * ( \k / (2*\np+2) ) + 90 }}
			\node[circle,line width=\lineW mm,BlobLineColour!80,draw,fill=BlobFillColour!10] (V\k) at (\angle: 1)[scale=\blobSc] {#2};
			\draw [line width=\lineW mm]  \foreach \x [remember=\x as \lastx (initially 1)] in {2,...,\np}{(V\lastx) to [bend right] (V\x)};
	}
	\draw [line width=\lineW mm] (V1) to [bend left] (0,1);
	\draw [line width=\lineW mm] (V\np) to [bend right] (0,-1);
}

\newcommand\rightArc[2]{
	\pgfmathsetmacro{\np}{#1} 
	\ifthenelse{\equal{\np}{1}}	{
		\def\angle{0}
		\node[circle,line width=\lineW mm,BlobLineColour!80,draw,fill=BlobFillColour!10] (V\np) at (\angle: 1)[scale=\blobSc] {#2};
	}	{
		\foreach \k in {1,...,\np}
			\def\angle{{ -1 * 360 * ( \k / (2*\np+2) ) + 90 }}
			\node[circle,line width=\lineW mm,BlobLineColour!80,draw,fill=BlobFillColour!10] (V\k) at (\angle: 1)[scale=\blobSc] {#2};
			\draw [line width=\lineW mm] \foreach \x [remember=\x as \lastx (initially 1)] in {2,...,\np}{(V\lastx) to [bend left] (V\x)};
	}
	\draw [line width=\lineW mm] (V1) to [bend right] (0,1);
	\draw [line width=\lineW mm] (V\np) to [bend left] (0,-1);
}

\newcommand\gluonLine[4]{
	\draw[decorate, decoration={snake,amplitude=.4mm,segment length=2mm,post length=0mm}, line width=\lineW mm, GluonColour!100] (#1,#2) -- (#3,#4);
}

\newcommand\ghostLine[5]{
	\tikzset{->-/.style={decoration={
  markings,
  mark=at position ##1 with {\arrow{>}}},postaction={decorate}}}
	\draw[->-=#5, loosely dotted, line width=\lineW mm, GhostColour!100]  (#1,#2) -- (#3,#4);
}

\newcommand\quarkLine[5]{
	\tikzset{->-/.style={decoration={
  markings,
  mark=at position ##1 with {\arrow{>}}},postaction={decorate}}}
	\draw[->-=#5, line width=\lineW mm, QuarkColour!100]  (#1,#2) -- (#3,#4);
} 

\newcommand\gluonLoop[3]{
	\draw[decorate, decoration={snake,amplitude=.4mm,segment length=2mm,post length=0mm}, line width=\lineW mm, GluonColour!100,fill=White] (#1,#2) circle (#3);
}

\newcommand\ghostLoop[5]{
	\tikzset{deco/.style n args={4}{decoration={markings, mark=at position ##1 with { \draw [<-] (0,0) --  (3pt,0)node [near end,##2=5 pt]{##3};}, mark=at position ##4 with { \draw [<-] (0,0) --  (3pt,0)node [near end,##2=5 pt]{##3};}},  postaction={decorate}}};
	\draw[deco={#4}{left}{}{#5},loosely dotted, line width=\lineW mm, GhostColour!100,->>,fill=White] (#1,#2) circle (#3);
}

\newcommand\quarkLoop[5]{
	\tikzset{deco/.style n args={4}{decoration={markings, mark=at position ##1 with { \draw [<-] (0,0) --  (3pt,0)node [near end,##2=5 pt]{##3};}, mark=at position ##4 with { \draw [<-] (0,0) --  (3pt,0)node [near end,##2=5 pt]{##3};}},  postaction={decorate}}};
	\draw[deco={#4}{left}{}{#5}, line width=\lineW mm, QuarkColour!100,->>,fill=White] (#1,#2) circle (#3);
}

\newcommand\doublePlainLoop[3]{
	\draw[style=double, line width=\lineW mm, DoublePlainColour!100,fill=White] (#1,#2) circle (#3);
}

\newcommand\gluonArc[5]{
	\draw[decorate, decoration={snake,amplitude=.4mm,segment length=2mm,post length=0mm}, line width=\lineW mm, GluonColour!100] ({#1+#3*cos(#4)},{#2+#3*sin(#4)}) arc (#4:#5:#3);
}

\DeclareMathOperator*{\sumint}{                                      \mathchoice
  {\ooalign{$\displaystyle\sum$\cr\hidewidth$\displaystyle\int$\hidewidth\cr}}
  {\ooalign{\raisebox{.14\height}{\scalebox{.7}{$\textstyle\sum$}}\cr\hidewidth$\textstyle\int$\hidewidth\cr}}
  {\ooalign{\raisebox{.2\height}{\scalebox{.6}{$\scriptstyle\sum$}}\cr$\scriptstyle\int$\cr}}
  {\ooalign{\raisebox{.2\height}{\scalebox{.6}{$\scriptstyle\sum$}}\cr$\scriptstyle\int$\cr}}
  }

  \newcommand{\RePart}[1]{\mathrm{Re}#1}
  \newcommand{\ImPart}[1]{\mathrm{Im}#1}

\allowdisplaybreaks

\begin{document}

\title{The chromoelectric adjoint correlators in Euclidean space at next-to-leading order}
\preprint{TUM-EFT 190/24}

\author[a,b,c]{Nora Brambilla,}
\author[a,d]{Panayiotis Panayiotou,}
\author[a,d,e,f]{Saga Säppi}
\author[a]{and Antonio Vairo}
\affiliation[a]{Technical University of Munich, TUM School of Natural Sciences, Physics Department,\\   
James-Franck-Str.~1, 85748 Garching, Germany}
\affiliation[b]{Technical University of Munich, Institute for Advanced Study, \\ 
Lichtenbergstrasse 2 a, 85748 Garching, Germany}
\affiliation[c]{Technical University of Munich, Munich Data Science Institute, \\ 
Walther-von-Dyck-Strasse 10, 85748 Garching, Germany}
\affiliation[d]{Excellence Cluster ORIGINS, \\
Boltzmannstrasse 2, 85748 Garching, Germany}
\affiliation[e]{Instituto de Ciencias del Espacio (ICE, CSIC),  \\
  C. Can Magrans s.n., 08193, Cerdanyola del Vallès, Catalonia, Spain}
\affiliation[f]{Institut d'Estudis Espacials de Catalunya (IEEC), \\
  08860, Castelldefels (Barcelona), Catalonia, Spain}
\emailAdd{nora.brambilla@tum.de}
\emailAdd{panayiotis.panayiotou@tum.de}
\emailAdd{sappi@ieec.cat}
\emailAdd{antonio.vairo@tum.de}

\abstract{
   The physics of quarkonium created in heavy-ion collisions is intrinsically connected to the correlation functions of adjoint chromoelectric fields in quantum chromodynamics. 
   We study such correlation functions in a weak-coupling expansion in a thermal medium. 
   We identify three distinct gauge-invariant correlators, and evaluate them to next-to-leading order.
   Two of the resulting correlators turn out to be asymmetric. 
   We pinpoint the source of this asymmetry to Matsubara zero modes associated with Wilson lines. 
   The results are shown to agree well with recent lattice calculations at high temperatures.
}

\maketitle
\newpage

\section{Introduction}
Heavy quarkonium is a key probe of the quark--gluon plasma (QGP), the high-temperature, deconfined, phase of quantum chromodynamics (QCD). 
Due to the short lifetime of the QGP produced in heavy-ion collisions, quarkonium suppression offers an indirect yet clean probe to study the conditions found in relativistic heavy-ion collisions and, by extension, in the early universe. 
This was originally proposed by Matsui and Satz \cite{Matsui:1986dk}, who suggested that in-medium colour-screening effects could reduce the yield of $J/\psi$, a charm--anticharm bound state,
in heavy-ion collisions compared to proton--proton collisions.

Relativistic heavy-ion experiments have provided a wealth of information on quarkonium suppression mechanisms\cite{PHENIX:2011img,ALICE:2012jsl,CMS:2016rpc,ATLAS:2018hqe,STAR:2019fge}.
These mechanisms are attributed to medium-induced effects such as plasma screening \cite{Karsch:1987pv},
dissociation\cite{Kharzeev:1994pz,Xu:1995eb,Grandchamp:2001pf,Grandchamp:2002wp,Laine:2006ns,Brambilla:2008cx,Brambilla:2011sg,Brambilla:2013dpa}
and regeneration \cite{Braun-Munzinger:2000csl,Thews:2000rj,Andronic:2007bi,Zhao:2007hh,Brambilla:2023hkw} of quarkonium. 
We are primarily interested in furthering our understanding of the propagation of heavy quarkonium in the QGP. 
The scales involved in heavy bound systems are well-separated, which calls for the use of non-relativistic effective field theories to simplify the description of the system \cite{Caswell:1985ui,Bodwin:1994jh,Brambilla:2004jw}. 
The typical hierarchy of scales in these systems is $M \gg 1/a_0\gg E$, where $M$ is the mass of the heavy particle, $1/a_0$ is the inverse of the Bohr radius and $E$ is the binding energy. 
Two additional scales are provided by the QGP itself, as it is the thermal medium in which the heavy bound state propagates. 
These scales are the temperature $T$ and the associated Debye screening mass $m_D \sim g_s T$. 
Here, $g_s$ is the strong coupling, which is small at large temperatures. 
A small value of the coupling also makes the thermal scales parametrically separated. 
Effective field theories are constructed by integrating out the separated energy scales in succession. 
For our case, the relevant effective field theory is potential non-relativistic quantum chromodynamics (pNRQCD) at finite temperature \cite{Brambilla:2008cx,Escobedo:2008sy,Brambilla:2010vq}. 

Recent studies \cite{Akamatsu:2011se,Akamatsu:2014qsa,Brambilla:2016wgg,Blaizot:2017ypk,Brambilla:2017zei,Yao:2018nmy,Blaizot:2018oev,Brambilla:2019tpt,Brambilla:2020qwo,Akamatsu:2020ypb,Yao:2020eqy,Miura:2022arv} use effective field theories,
such as pNRQCD, to model the propagation of quarkonium in a medium as an open quantum system, with the QGP acting as the environment. 
In this framework, the evolution of the quarkonium density matrix in the QGP is governed by a master equation that depends on some transport coefficients. 
It was first established in \cite{Brambilla:2016wgg} that these transport coefficients could be extracted from the integrals of chromoelectric correlators connected by \emph{adjoint} temporal Wilson lines. 
In the open quantum system framework, this emerges by tracing out the environmental degrees of freedom.\footnote{
Similar correlators also appear in studies of dark matter\cite{Binder:2021otw,Biondini:2023zcz}.} 
These chromoelectric correlators are the subject of this work. 

Previous works on the correlators \cite{Binder:2021otw,Eller:2019spw} have focussed on the real-time formulation of thermal QCD and/or dealt with the extraction of transport coefficients related to the propagation of a heavy quark in the plasma. 
Recently, the analytic continuation of one of the real-time correlators responsible for the singlet--octet transition in pNRQCD has been performed \cite{Scheihing-Hitschfeld:2023tuz}.
Here we use the imaginary-time formulation of perturbative QCD, and evaluate the correlators directly at weak coupling, instead of computing the spectral function. 
Our choice makes it easier to make contact with lattice QCD studies. 
Indeed, we are able to provide a perturbative explanation for the recent findings of the first lattice calculation 
of these correlators \cite{Brambilla:2025cqy} showing, in particular,  that the correlator responsible for singlet--octet dipole transition exhibits asymmetry on the thermal circle. 

While not directly relevant to quarkonium, one should also keep in mind the well-established framework of heavy-quark diffusion valid for a weakly-coupled plasma with the hierarchy $M \gg T \gg m_D$.
Analogously to pNRQCD and quarkonium, heavy-quark effective theory is used to study the diffusion of heavy quarks in a thermal medium. 
Within this approach, the authors of \cite{Caron-Huot:2009ncn} derived an expression for the heavy quark diffusion coefficient in the QGP in terms of a gauge invariant correlator of chromoelectric fields connected by \emph{fundamental} Wilson lines
within the imaginary-time formalism.
In \cite{Burnier:2010rp}, the spectral function of the fundamental chromoelectric correlator was evaluated and subsequently a prediction for the heavy quark diffusion coefficient was made. 
This correlator containing fundamental Wilson lines coincides, up to Casimir scaling, with one of the three adjoint correlators considered in this work.

The paper is structured as follows. 
First, we give precise definitions of the three gauge-invariant chromoelectric correlators that can be constructed from two chromoelectric fields and Wilson lines in the adjoint representation, and discuss their distinct physical meanings. 
Next, we evaluate the adjoint correlators up to next-to-leading order (NLO). 
We use modern integration-by-parts methodology to factorise the vast majority of the integrals into one-loop integrals with known closed forms.
Lastly, we show our results and discuss their implications in view of recent lattice findings.
The work is completed by a set of more technical appendices.

\section{The chromoelectric correlators in the adjoint representation}
We define three distinct chromoelectric correlators in the adjoint representation: the \emph{upper} correlator ($U$), the \emph{lower} correlator ($L$), and the \emph{symmetric} correlator ($S$), which are defined as 
\begin{align}
\left\langle EE \right\rangle_U \equiv & -\left\langle g_s E^a_{i}(0) U^{ab}(0,t) g_s E^b_{i} (t) \right\rangle T^{-4},\label{defEEU}\\
\left\langle EE \right\rangle_L \equiv & -\left\langle g_s E^b_{i}(0) g_s E^a_{i} (t) U^{ab}(t,1/T) \right\rangle T^{-4}, \label{defEEL} \\
\left\langle EE \right\rangle_S \equiv & -\left\langle g_s E_i^{ab}(0) U^{bc}(0,t)g_s E_i^{cd}(t) U^{da}(t,1/T) \right\rangle T^{-4}, \label{defEES}
\end{align}
respectively. 
We have divided the correlators by $T^4$ in order to make them dimensionless. 
Above, $E_i^a (t)$ are the components of the chromoelectric field, $U^{ab} (t,t')$ are the components of a Wilson line in the adjoint representation connecting points on the Euclidean thermal circle $(0,1/T)$
at times $t$ and $t'$, and $E^{ab}_{i}(t)\equiv x_{abc} E^c_i (t)$ with $x_{abc} \in \{f_{abc},d_{abc}\}$, $(f_{abc})\,d_{abc}$ being the (anti-)symmetric structure constants of $\mathrm{SU}(N_c)$. 
The angular brackets stand for thermal field theory averages. 
For precise definitions of the Wilson line and the chromoelectric field, see Appendix \ref{app:conventions}. 
Here, we only consider quantities at the spatial origin, and because of that we suppress the dependence on the spatial coordinates. 

The three correlators have distinct interpretations. 
The upper correlator is a widely-studied quantity associated with quarkonium dissociation in pNRQCD \cite{Brambilla:2016wgg,Brambilla:2017zei,Binder:2021otw},
while the lower correlator, or rather its real-time counterpart, is associated with recombination \cite{Yao:2020eqy}. 
In Euclidean space, the two are related by $\langle EE \rangle_U (t) = \langle EE\rangle_L (1/T-t)$. 
The symmetric correlator with antisymmetric structure constants has no known simple physical application but appears to be related to the diffusion of adjoint heavy particles, 
whereas the symmetric correlator with symmetric structure constants appears to be related to the octet-to-octet dipole transition in pNRQCD \cite{Brambilla:QuarkoniumTransportCoefficients}.
The symmetric correlator is manifestly symmetric on the thermal circle.
This makes it attractive from the point of view of lattice field theory, where the symmetry property allows for solving the inverse problem of obtaining a spectral function corresponding to the correlator \cite{Brambilla:2022xbd}.
Lattice simulations suggest that the upper and lower correlators ought to be asymmetric, and in this paper we find an explicit, closed-form expression for the leading-order (LO) asymmetry of the correlators. 
We discuss the matter of (a)symmetry in more detail in sections \ref{sec:resultsasy} and \ref{sec:discussion}.

In existing literature, the most commonly studied Euclidean correlator is  \cite{Caron-Huot:2009ncn,Burnier:2010rp}
\begin{equation}
\langle EE \rangle_F \equiv -\left\langle \mathrm{Tr}\left[ g_s E_i (0) U(0,t) g_s E_i(t) U(t,1/T) \right] \right\rangle T^{-4},
\label{eq:EEF}
\end{equation}
where the fields $E_i$ and Wilson lines $U$ are now in the fundamental representation, and the trace acts on the $N_c \times N_c$ generators of $\mathrm{SU}(N_c)$. 
This correlator is related to heavy quark diffusion rather than quarkonium diffusion. 
In \cite{Burnier:2010rp}, the spectral function associated with the correlator in eq. \eqref{eq:EEF}  was calculated at NLO. 
Here in section \ref{sec:resultsS}, we give, to the best of our knowledge for the first time, a direct evaluation of the fundamental correlator at NLO. 
The fundamental correlator provides a useful cross-check for us, as similar methods are used for correlators in both the fundamental and adjoint representations. 

Along with the abundance of imaginary-time integration methods, lattice field theory serves as our motivation for remaining in Euclidean space, as it allows for a simple, unambiguous,
and direct comparison of the correlators with lattice results. 
In particular, the three correlators defined above are the \emph{only} gauge-invariant correlators made up of two chromoelectric fields and Wilson lines connecting them. 
In contrast, one can, in principle, define infinitely many similar correlators in the real-time formalism, as there are infinitely many inequivalent paths one can draw to connect the two electric fields in the complexified time plane. 
The Minkowskian correlators have attracted significant interest in recent years, with some of them considered at NLO in \cite{Binder:2021otw}.

\section{Evaluating the correlators}

\subsection{Diagrammatics}

The computation of the correlators is most conveniently started with diagrammatics: at NLO, there are nine distinct topologies contributing to the upper and lower correlators, which we list below divided into subcategories. 

The first diagrams contain gluon self-energy contributions: 
\begin{equation}\label{gluonselfenergy}
\raisebox{-0.44\height}{
\begin{tikzpicture}[scale=\picSc]
	\doublePlainLoop{2}{0}{2}
	\gluonLine{0}{0}{4}{0}
	\blobNode{0}{0}{\bigBlobSc}{$E$}
	\blobNode{4}{0}{\bigBlobSc}{$E$},
	\gluonLoop{2}{-0.1}{0.8}
\end{tikzpicture}}  \quad
\raisebox{-0.44\height}{
\begin{tikzpicture}[scale=\picSc]
	\doublePlainLoop{2}{0}{2}
	\gluonLine{0}{0}{4}{0}
	\blobNode{0}{0}{\bigBlobSc}{$E$}
	\blobNode{4}{0}{\bigBlobSc}{$E$},
	\gluonLoop{2}{+0.7}{0.8}
\end{tikzpicture}} \quad
\raisebox{-0.44\height}{
\begin{tikzpicture}[scale=\picSc]
	\doublePlainLoop{2}{0}{2}
	\gluonLine{0}{0}{4}{0}
	\blobNode{0}{0}{\bigBlobSc}{$E$}
	\blobNode{4}{0}{\bigBlobSc}{$E$},
	\ghostLoop{2}{-0.1}{0.8}{0.25}{0.75}
\end{tikzpicture}}  \quad
\raisebox{-0.44\height}{
\begin{tikzpicture}[scale=\picSc]
	\doublePlainLoop{2}{0}{2}
	\gluonLine{0}{0}{4}{0}
	\blobNode{0}{0}{\bigBlobSc}{$E$}
	\blobNode{4}{0}{\bigBlobSc}{$E$},
	\quarkLoop{2}{-0.1}{0.8}{0.25}{0.75}
\end{tikzpicture}}\,\,.
\end{equation}
We use plain double lines to denote the Wilson lines and blobs to denote the chromoelectric field insertions (with one of them implicitly at Euclidean time $0$ and the other at $t$),
along with a common choice of notation for the QCD propagators (see Appendix \ref{app:conventions}).
While the graphical notation is strictly speaking only valid for correlators with two Wilson lines, diagrammatically an adjoint Wilson line without internal lines connecting to it is equivalent to a colour-space $\delta_{ab}$,
and as such we may implicitly omit the lower (upper) line when treating the upper (lower) correlator. 
This allows us to use uniform diagrammatic notation for all correlator types. 

In addition to the above ones, there are two more topologies that contain no contributions from the Wilson lines:
\begin{equation}\label{top1}
\raisebox{-0.44\height}{
\begin{tikzpicture}[scale=\picSc]
	\doublePlainLoop{2}{0}{2}
	\gluonArc{2}{-2.25}{3}{40}{140}
	\gluonArc{2}{+2.25}{3}{-40}{-140}
	\blobNode{0}{0}{\bigBlobSc}{$E$}
	\blobNode{4}{0}{\bigBlobSc}{$E$} 
\end{tikzpicture}}\quad
\raisebox{-0.44\height}{
\begin{tikzpicture}[scale=\picSc]
	\doublePlainLoop{2}{0}{2}
	\gluonLoop{1.3}{-0.1}{0.8}
	\gluonLine{2.1}{0}{4}{0}
	\blobNode{0}{0}{\bigBlobSc}{$E$}
	\blobNode{4}{0}{\bigBlobSc}{$E$}
\end{tikzpicture}} \quad
\raisebox{-0.44\height}{
\begin{tikzpicture}[scale=\picSc]
	\doublePlainLoop{2}{0}{2}
	\gluonLoop{2.7}{-0.1}{0.8}
	\gluonLine{0}{0}{1.9}{0}
	\blobNode{0}{0}{\bigBlobSc}{$E$}
	\blobNode{4}{0}{\bigBlobSc}{$E$}
\end{tikzpicture}}\,\,.
\end{equation}
There are also three diagrams in which Wilson lines are expanded to first order:
\begin{equation}\label{top2}
\raisebox{-0.44\height}{
\begin{tikzpicture}[scale=\picSc]
	\doublePlainLoop{2}{0}{2}
	\gluonLine{0}{0}{4}{0}
	\gluonArc{0}{2}{1.5}{-5}{-90}
	\blobNode{0}{0}{\bigBlobSc}{$E$}
	\blobNode{4}{0}{\bigBlobSc}{$E$} 
\end{tikzpicture}} \quad
\raisebox{-0.44\height}{
\begin{tikzpicture}[scale=\picSc]
	\doublePlainLoop{2}{0}{2}
	\gluonLine{0}{0}{4}{0}
	\gluonArc{4}{2}{1.6}{185}{270}
	\blobNode{0}{0}{\bigBlobSc}{$E$}
	\blobNode{4}{0}{\bigBlobSc}{$E$} 
\end{tikzpicture}} \quad 
\raisebox{-0.44\height}{
\begin{tikzpicture}[scale=\picSc]
	\doublePlainLoop{2}{0}{2}
	\gluonLine{0}{0}{4}{0}
	\gluonLine{2}{1.9}{2}{0}
	\blobNode{0}{0}{\bigBlobSc}{$E$}
	\blobNode{4}{0}{\bigBlobSc}{$E$}
\end{tikzpicture}}\,\,,
\end{equation}
and likewise three in which they are expanded to second order: 
\begin{equation}\label{top3}
\raisebox{-0.44\height}{
\begin{tikzpicture}[scale=\picSc]
	\doublePlainLoop{2}{0}{2}
	\gluonArc{0}{2}{1.5}{-5}{-90}
	\gluonArc{4}{2}{1.5}{185}{270}
	\blobNode{0}{0}{\bigBlobSc}{$E$}
	\blobNode{4}{0}{\bigBlobSc}{$E$}
\end{tikzpicture}} \quad 
\raisebox{-0.44\height}{
\begin{tikzpicture}[scale=\picSc]
	\doublePlainLoop{2}{0}{2}
	\gluonArc{0}{3}{3}{-25}{-42}
	\gluonArc{0}{3}{3}{-52}{-90}
	\gluonArc{4}{3}{3}{205}{270}
	\blobNode{0}{0}{\bigBlobSc}{$E$}
	\blobNode{4}{0}{\bigBlobSc}{$E$}
\end{tikzpicture}} \quad
\raisebox{-0.44\height}{
\begin{tikzpicture}[scale=\picSc]
	\doublePlainLoop{2}{0}{2}
	\gluonLine{0}{0}{4}{0}
	\gluonLine{0.5}{1}{3.5}{1}
	\blobNode{0}{0}{\bigBlobSc}{$E$}
	\blobNode{4}{0}{\bigBlobSc}{$E$}
\end{tikzpicture}}\,\,.
\end{equation}
We have only drawn the diagrams for the upper correlator here, but the corresponding contributions to the other correlator types are obtained as trivial extensions. 

Finally, there are two additional topologies which only appear if one includes two distinct Wilson lines, i.e. for the symmetric (and fundamental) correlator(s): 
\begin{equation}\label{top4}
\raisebox{-0.44\height}{
\begin{tikzpicture}[scale=\picSc]
	\doublePlainLoop{2}{0}{2}
	\gluonArc{0}{2}{2}{-5}{-90}
	\gluonArc{4}{-2}{2}{90}{175}
	\blobNode{0}{0}{\bigBlobSc}{$E$}
	\blobNode{4}{0}{\bigBlobSc}{$E$},
\end{tikzpicture}} \quad
\raisebox{-0.44\height}{
\begin{tikzpicture}[scale=\picSc]
	\doublePlainLoop{2}{0}{2}
	\gluonArc{0}{-2}{2}{0}{90}
	\gluonArc{4}{+2}{2}{180}{270}
	\blobNode{0}{0}{\bigBlobSc}{$E$}
	\blobNode{4}{0}{\bigBlobSc}{$E$},
\end{tikzpicture}} \quad
\raisebox{-0.44\height}{
\begin{tikzpicture}[scale=\picSc]
	\doublePlainLoop{2}{0}{2}
	\gluonLine{0}{0}{4}{0}
	\gluonLine{2}{1.9}{2}{0.3}
	\gluonLine{2}{-0.3}{2}{-1.9}
	\blobNode{0}{0}{\bigBlobSc}{$E$}
	\blobNode{4}{0}{\bigBlobSc}{$E$}
\end{tikzpicture}} \,\,.
\end{equation}

\subsection{Evaluating sum-integrals}\label{sec:SumInt}
To establish notation (further elaborated in Appendix \ref{app:conventions}), we evaluate the LO correlator. 
At LO, the upper and lower correlators are trivially the same, and the fundamental and symmetric correlators differ from them simply by an overall group factor. 
For the upper and lower correlators, 
\begin{align}\label{eq:LOEE}
  \langle  EE\rangle_{\mathrm{LO}} &= - \delta_{ab}\left\langle g_s E_i^a(0) g_s E_i^b(t) \right\rangle T^{-4}= 
- g_s^2 T^{-4}\raisebox{-0.44\height}{
\begin{tikzpicture}[scale=0.75*\picSc]
	\doublePlainLoop{2}{0}{2}
	\gluonLine{0}{0}{4}{0}
	\blobNode{0}{0}{\bigBlobSc}{$E$}
	\blobNode{4}{0}{\bigBlobSc}{$E$},
\end{tikzpicture}} 
\nonumber \\
 & =
- g_s^2 T^{-4} \Bigg[\left\langle \wick{\partial_0 \c1 A_i^a(0) \partial_0 \c1 A_i^a(t)}\right\rangle -\left\langle \wick{\partial_i \c1 A_0^a(0) \partial_0 \c1 A_i^a(t)}\right\rangle \nonumber \\
&\qquad\qquad -\left\langle \wick{\partial_0 \c1 A_i^a(0) \partial_i \c1 A_0^a(t)}\right\rangle+\left\langle \wick{\partial_i \c1 A_0^a(0) \partial_i \c1 A_0^a(t)}\right\rangle  \Bigg]
\nonumber \\
& = - g_s^2 T^{-4}  \delta_a^a \sumint_P e^{ip_0 t}\left[ p_0^2 G_{ii}(P) + p^2 G_{00}(P) -2 p_0 p_i G_{0i}(P) \right] 
\nonumber \\ 
 &= - d_A g_s^2 T^{-4} (d-1) \sumint_P \frac{p_0^2}{P^2}e^{ip_0 t}
 \nonumber\\
 &=  d_A g_s^2  \pi^2 \left[\mathrm{cos}(2\pi Tt)+2\right]\mathrm{csc}^4(\pi Tt) +O(\varepsilon),
\end{align}
where we have expanded the dimensionally regulated integral up to order $\varepsilon$ corrections; the exact expression of it in $d$ dimensions can be found below.

The computation of the two-loop diagrams proceeds in a similar fashion. 
We write down the diagrams using Feynman rules and perform colour and Lorentz contractions to scalarise the diagrams. 
Already at this point, a large number of terms vanish by rotational invariance, as they factor into rank-one integrals of the form $\displaystyle \sumint_{P} p^i f(p_0,p) = 0$ for some scalar-valued function $f$.  
Following this, we perform any integrals over the auxiliary time variables appearing in the Wilson lines, unless otherwise stated. 
For bosonic contributions, the result of this procedure will be a linear combination of sum-integrals of the form 
\begin{equation} \label{eq:bosonicm}
\mathcal{M}^{abc}_{nml} (t,t') \equiv \sumint_{PQ} \frac{p_0^a q_0^b r_0^c}{P^{2n}Q^{2m}R^{2l}} e^{i p_0 t}e^{i q_0 t'}; \quad R\equiv P+Q;\quad\quad a,b,c,n,m,l\in\mathbb{Z}_{\geq0}.
\end{equation}
In a typical case, a subset of the indices $\lbrace a,b,c \rbrace$ and/or $\lbrace n,m,l \rbrace$ vanishes, and one can easily reduce the set of scalarised integrals to a much smaller subset,
often factorising them into products of one-loop integrals in the process.
However, for two-loop sum-integrals a much stronger result exists: the so-called integration-by-parts results applied to the spatial part of the sum-integral guarantee that,
for example, \emph{all} sum-integrals $\mathcal{M}^{abc}_{nml} (t,t')$ for which $\lbrace a,b,c\rbrace\in \mathbb{Z}_+^3$ 
can be factorised into one-loop integrals \cite{Davydychev:2022dcw,Davydychev:2023jto}.
Furthermore,
the generic one-loop sum-integral can be evaluated analytically:
\begin{align} \label{eq:eresult}
  \mathcal{E}^{a}_{m}(t) \equiv \sumint_P e^{ip_0 t} \frac{p_0^{a}}{P^{2m}}=& \frac{T^{4-2m+a}}{\left(2\pi\right)^{2m-a}}
  \frac{\Gamma\left(m-\frac{d}{2}\right)}{\Gamma\left(m\right)}\pi^{\frac{3}{2}}\left(\frac{e^{\gamma_{E}}\bar{\Lambda}^{2}}{4\pi^{2}T^{2}}\right)^{\frac{3-d}{2}}\nonumber\\
    &\times\left[\mathrm{Li}_{2m-a-d}\left(e^{2\pi iTt}\right)+(-1)^{a}\mathrm{Li}_{2m-a-d}\left(e^{-2\pi iTt}\right)\right],
    \end{align}
where $\mathrm{Li}_s(z)$ is the standard polylogarithm.
For details, see Appendix  \ref{app:ixr}.
In the case at hand, the only nontrivial integration-by-parts identity required is
\begin{align}
\int_{\mathbf{pq}} &\frac{1}{(p_0^2+p^2)(q_0^2+q^2)\left[(p_0+q_0)^2+(\mathbf{p}+\mathbf{q})^2\right]} \nonumber\\
&= -\frac{d-2}{2\left(d-3\right)}
\Bigg[\frac{I_{1}\left(p_{0}\right)I_{1}\left(p_{0}+q_{0}\right)}{p_{0}\left(p_{0}+q_{0}\right)}+\frac{I_{1}\left(q_{0}\right)I_{1}\left(p_{0}+q_{0}\right)}{q_{0}\left(p_{0}+q_{0}\right)}
  -\frac{I_{1}\left(p_{0}\right)I_{1}\left(q_{0}\right)}{p_{0}q_{0}}\Bigg], 
\end{align}
in which a standard textbook integral, 
\begin{equation}
I_{n}\left(m\right)\equiv\int_{\boldsymbol{p}}\frac{1}{\left(\boldsymbol{p}^{2}+m^{2}\right)^n}, 
\end{equation}
appears.
The unique fermionic contribution comes from the last diagram in \eqref{gluonselfenergy}.
Its evaluation proceeds in an equivalent way, except that the sum-integrals are of the general form 
\begin{equation}
\sumint_{\lbrace{PQ\rbrace}} \frac{p_0^a q_0^b r_0^c}{P^{2n}Q^{2m}R^{2l}} e^{i p_0 t}e^{i q_0 t'}; \quad R\equiv P+Q, 
\end{equation}
where curly brackets indicate loop momenta with fermionic Matsubara modes. 
The fermionic one-loop integral can be written in terms of the bosonic ones via the usual relation connecting fermionic and bosonic Matsubara sums
\begin{equation} \label{eq:1lfermion}
     \sumint_{\lbrace P\rbrace} e^{ip_0 t} \frac{p_0^{a}}{P^{2m}}=2^{2m-a-d}\mathcal{E}^a_m \left( t/2 \right)-\mathcal{E}^a_m \left( t \right).
\end{equation}

\subsection{Nonfactorisable contribution}\label{nonfact}
There is exactly one nontrivial integral that cannot be factorised in the way described above or otherwise computed analytically. 
This contribution is contained in a single topology, the one where a Wilson line is connected to a three-point vertex. 
Taking the upper correlator as an example, it can be written in terms of scalar integrals as
\begin{align}
\raisebox{-0.44\height}{
\begin{tikzpicture}[scale=0.75*\picSc]
	\doublePlainLoop{2}{0}{2}
	\gluonLine{0}{0}{4}{0}
	\gluonLine{2}{1.9}{2}{0}
	\blobNode{0}{0}{\bigBlobSc}{$E$}
	\blobNode{4}{0}{\bigBlobSc}{$E$}
\end{tikzpicture}} \quad \Bigg|_\mathrm{non-factorisable} & \propto  \quad
\frac{i}{2}\int_{0}^{t}\mathrm{d}t'\sumint_{PQ}\frac{e^{ip_{0}\left(t-t'\right)}q_{0}e^{iq_{0}t}}{\left(\left(\mathbf{p}+\mathbf{q}\right)^{2}+(p_{0}+q_0)^{2}\right)\left(q^{2}+q_{0}^{2}\right)} \nonumber\\
&=\frac{i}{2}\int_{0}^{t}\mathrm{d}t'\mathcal{E}_{1}^{1}\left(t'\right)\mathcal{E}_{1}^{0}\left(t-t'\right), \label{eq:Ixr}
\end{align}
where we have omitted amongst others the overall factor of $g_s^4d_AN_c$ and $t'$ is the time coordinate of the gauge field appearing in the Wilson line.

From eq. \eqref{eq:Ixr}, it follows that the time integral is proportional to $iq^0[e^{iq_0t} - e^{i(p_0+q_0)t}]/p_0$ if $p_0 = 2\pi n T$ with $n\in \mathbb{Z}\backslash\lbrace 0 \rbrace$ and to $t q_0 e^{iq_0t}$ if $p_0=0$;
note that the integral vanishes if $q_0=0$.
Hence, the zero-mode contributions have quite a different time-dependence from non-zero-mode contributions. 
The latter is symmetric under  $t \to 1/T-t$, whereas the zero-mode contribution is not.
In particular, the zero-mode contribution associated with the Wilson line probes the length of the Wilson line, 
which is a primary distinguishing feature between the upper and lower correlators. 

To explicitly separate the symmetric from the non-symmetric contribution, it is useful to split the integral \eqref{eq:Ixr} into
\begin{align}
    \frac{i}{2}&\int_{0}^{t}\mathrm{d}t'\sumint_{PQ}\frac{e^{ip_{0}\left(t-t'\right)}q_{0}e^{iq_{0}t}}{\left(\left(\mathbf{p}+\mathbf{q}\right)^{2}+(p_{0}+q_0)^{2}\right)\left(q^{2}+q_{0}^{2}\right)} \nonumber\\
    &= \frac{i}{2}\int_{0}^{t}\mathrm{d}t'\mathcal{E}_{1}^{1}\left(t'\right)\mathcal{E}_{1}^{0}\left(t-t'\right)|_{\mathrm{p_0 \neq 0}}
    +\frac{t}{2}iT\int_{\mathbf{p}}\sumint_{Q}\frac{q_{0}e^{iq_{0}t}}{\left(\left(\mathbf{p}+\mathbf{q}\right)^{2}+q_0^{2}\right)\left(q^{2}+q_{0}^{2}\right)},
    \label{eq:Ixr2}
\end{align}
where the first term does not contain zero modes and is symmetric under $t \to 1/T-t$, while the second term is the non-symmetric zero-mode contribution, proportional to the length of the Wilson line, which in this case is $t$.
Hence, this contribution changes for the other two correlators.
We give the explicit results for these integrals below.
We discuss the matter of asymmetry further in section \ref{sec:resultsasy}.  

We note that the $t'$-integral above does not appear to admit a closed-form solution. 
As the integral is ultraviolet divergent, it is convenient to first extract its divergent part before evaluating it numerically.
We discuss this procedure, and the numerical results, in detail in Appendix \ref{app:ixr}.

 \subsection{Zero-mode contributions}\label{sec:zeromode}
In two-loop sum-integrals, zero-mode contributions are generally of the form 
\begin{equation}
\mathcal{Z}^{a}_{nml} (t) \equiv \sumint_{P} \int_{\mathbf{q}} \frac{p_0^a e^{i p_0 t}}{(\mathbf{p}+\mathbf{q})^{2n}(p_0^2+p^2)^{m}(p_0^2+q^2)^{l}}, 
\end{equation}
with $a,n,m,l\in\mathbb{Z}_{\ge0}$.
Note that the zero-mode contribution to the above integral  (i.e. the "doubly-zero" mode of the full two-loop sum-integral) vanishes as a scaleless integral
 in dimensional regularisation.
Various recursion and factorisation results can be derived for $\mathcal{Z}^{a}_{nml}(t)$, analogously to $\mathcal{M}^{abc}_{nml} (t,t')$, with the only nontrivial result needed here being the formula 
\begin{equation}\label{eq:zIBP}
\int_{\mathbf{pq}} \frac{1}{(\mathbf{p}+\mathbf{q})^2(p_0^2+p^2)(p_0^2+q^2)} = -\frac{d-2}{2p_0^2(d-3)}\int_{\mathbf{p}} \frac{1}{p_0^2+p^2}\int_{\mathbf{q}}\frac{1}{p_0^2+q^2}.
\end{equation}
With the help of integration-by-parts results, we find that the unique non-vanishing zero-mode contribution is the one responsible of the asymmetry of the upper and lower correlators over the thermal circle discussed in the previous section.
From eq. \eqref{eq:Ixr2}, it follows that we can express that zero-mode contribution as the sum-integral $\mathcal{Z}^1_{011}(t)$, after performing the spatial momentum translation 
$\mathbf{p}+\mathbf{q} \to \mathbf{p}$.
The sum-integral $\mathcal{Z}^1_{011}(t)$ can be evaluated analytically, giving

\begin{align} 
\mathcal{Z}^{1}_{011}(t) &= \frac{T^4}{8}\Gamma^2\left(1-\frac{d}{2}\right)\left(\frac{e^{\gamma_{E}}\bar{\Lambda}^{2}}{4\pi^{2}T^{2}}\right)^{3-d}\left[\mathrm{Li}_{3-2d}\left(e^{2\pi i T t}\right)-\mathrm{Li}_{3-2d}\left(e^{-2\pi i T t}\right)\right]\nonumber\\
&=2\pi T^4\left[\mathrm{Li}^{(1)}_{-3}(e^{2i\pi Tt})-\mathrm{Li}^{(1)}_{-3}(e^{-2i\pi Tt})\right]\varepsilon+O(\varepsilon^2)\nonumber\\
&=-\frac{3i}{\left(2\pi\right)^2}T^4\left[\zeta\left(4,Tt\right)-\zeta\left(4,1-Tt\right)\right]\varepsilon+O(\varepsilon^2),
\label{eq:zresult}
\end{align}
where $\mathrm{Li}^{(1)}_s(z)$ is the first derivative of the polylogarithm $\mathrm{Li}_s(z)$ with respect to $s$. To go from the second to the third equality we have used the relation $\mathrm{Li}_{s}(e^{2i\pi Tt})+(i)^{2s}\mathrm{Li}_{s}(e^{-2i\pi Tt}) = (2\pi)^si^s\zeta(1-s,Tt)/\Gamma(s)$, where $\zeta\left(s,a\right)$ is the Hurwitz zeta function.

\subsection{Results for the individual diagrams}\label{sec:diagrams}
We have implemented the computation in a symbolic code (making use of the FeynCalc package \cite{Shtabovenko:2020gxv,Shtabovenko:2016sxi,Mertig:1990an} to perform tensor contractions), 
and independently cross-checked the results by hand. 
At this order, computations by hand are still feasible, but the symbolic code provides a valuable proof-of-concept for proceeding to higher orders. 
We adopt the $R_\xi$ gauge, so that individual diagrams may depend on the gauge parameter $\xi$.

Below, we first list the results for each diagram contributing to the upper correlator, with an overall factor of $g_s^4 d_A N_c$ ($g_s^4 d_A N_f$ for the quark loop diagram \eqref{EEfermion}, $N_f$ being the number of massless quarks) omitted. 
The results are given in terms of products of the one-loop thermal scalar integrals $\mathcal{E}^{a}_{m}(t)$ (reduced to an independent subset), and the one convolution integral that contains the zero mode $\mathcal{Z}^{1}_{011}(t)$. 
We combine pairs of diagrams related by reflection about the vertical axis.
The results read
\begin{align}
\raisebox{-0.44\height}{
\begin{tikzpicture}[scale=0.75*\picSc]
	\doublePlainLoop{2}{0}{2}
	\gluonLine{0}{0}{4}{0}
	\blobNode{0}{0}{\bigBlobSc}{$E$}
	\blobNode{4}{0}{\bigBlobSc}{$E$},
	\gluonLoop{2}{-0.1}{0.8}
\end{tikzpicture}}  &\quad = \quad -\left[d+\frac{3}{2}+\frac{3}{2(d-3)}\right]\left[\mathcal{E}^0_1 (t)\right]^2\nonumber\\
&+\left[\frac{d^2}{2}-\frac{7d}{2}-1-\frac{5}{d-3}\right]\mathcal{E}^0_1 (t)\mathcal{E}^0_1 (0)+4\left[1+\frac{2}{d-3}\right]\mathcal{E}^1_2 (t)\mathcal{E}^1_1 (t)\nonumber \\
&+\Bigg \lbrace \left[\frac{1}{d-3}+1-\frac{d}{2}(d-3) \right]\left[\mathcal{E}^0_1 (t)\right]^2+\frac{d}{2}\mathcal{E}^0_1 (t)\mathcal{E}^0_1 (0)\nonumber \\
&+2\left[d-1-\frac{1}{d-3}\right]\mathcal{E}^1_1(t) \mathcal{E}^1_2(t)+\left(d-1\right)\left[\mathcal{E}^0_2(t)-\mathcal{E}^0_2 (0)\right]\mathcal{E}^{2}_{1}(t) \Bigg  \rbrace(1-\xi)\nonumber \\
&+\frac{d}{4}(d-2)\left[\mathcal{E}^0_1 (t)\right]^2 (1-\xi)^2,\label{eq:selfE} \\
\raisebox{-0.44\height}{
\begin{tikzpicture}[scale=0.75*\picSc]
	\doublePlainLoop{2}{0}{2}
	\gluonLine{0}{0}{4}{0}
	\blobNode{0}{0}{\bigBlobSc}{$E$}
	\blobNode{4}{0}{\bigBlobSc}{$E$},
	\gluonLoop{2}{+0.7}{0.8}
\end{tikzpicture}}  \quad &= \quad -\frac{d}{2}\left[d(d-3)+(1-\xi)\right]\mathcal{E}^0_1(t)\mathcal{E}^0_1(0), \\
\raisebox{-0.44\height}{
\begin{tikzpicture}[scale=0.75*\picSc]
	\doublePlainLoop{2}{0}{2}
	\gluonLine{0}{0}{4}{0}
	\blobNode{0}{0}{\bigBlobSc}{$E$}
	\blobNode{4}{0}{\bigBlobSc}{$E$},
	\ghostLoop{2}{-0.1}{0.8}{0.25}{0.75}
\end{tikzpicture}}   \quad &= \quad -\frac{1}{2}\left[1+\frac{1}{d-3}\right] \left[\mathcal{E}^{0}_{1}(t)-2\mathcal{E}^{0}_{1}(0)\right] \mathcal{E}^{0}_{1}(t), \\
\raisebox{-0.44\height}{
\begin{tikzpicture}[scale=0.75*\picSc]
	\doublePlainLoop{2}{0}{2}
	\gluonLine{0}{0}{4}{0}
	\blobNode{0}{0}{\bigBlobSc}{$E$}
	\blobNode{4}{0}{\bigBlobSc}{$E$},
	\quarkLoop{2}{-0.1}{0.8}{0.25}{0.75}
\end{tikzpicture}}  \quad & \quad= (d-1)\left\lbrace\left[\mathcal{E}^0_1 \left(t\right)\right]^2+4^{2-d}\left[\mathcal{E}^0_1 \left(\frac{t}{2}\right)\right]^2-2^{3-d}\mathcal{E}^0_1 \left(\frac{t}{2}\right)\mathcal{E}^0_1 \left(t\right)\right\rbrace\nonumber\\
&+\left(2^{2-d}-1\right)\left[(d-3)d+4+\frac{4}{d-3}\right]\mathcal{E}^0_1 \left(t\right)\mathcal{E}^0_1 \left(0\right)\nonumber \nonumber \\
&-\left[1+\frac{2}{d-3}\right]\left[2\mathcal{E}^1_1(t)-2^{2-d}\mathcal{E}^1_1\left(\frac{t}{2}\right)\right]\left[\mathcal{E}^1_2(t)-2^{3-d}\mathcal{E}^1_2\left(\frac{t}{2}\right)\right] , \label{EEfermion}\\
\raisebox{-0.44\height}{
\begin{tikzpicture}[scale=0.75*\picSc]
	\doublePlainLoop{2}{0}{2}
	\gluonArc{2}{-2.25}{3}{40}{140}
	\gluonArc{2}{+2.25}{3}{-40}{-140}
	\blobNode{0}{0}{\bigBlobSc}{$E$}
	\blobNode{4}{0}{\bigBlobSc}{$E$} 
\end{tikzpicture}}   \quad &= \quad -d\left[1+\frac{d-3}{2}(1-\xi)-\frac{d-2}{4}(1-\xi)^2\right]\left[\mathcal{E}^0_1 \left(t\right)\right]^2, \\
\raisebox{-0.44\height}{
\begin{tikzpicture}[scale=0.75*\picSc]
	\doublePlainLoop{2}{0}{2}
	\gluonLoop{1.3}{-0.1}{0.8}
	\gluonLine{2.1}{0}{4}{0}
	\blobNode{0}{0}{\bigBlobSc}{$E$}
	\blobNode{4}{0}{\bigBlobSc}{$E$}
\end{tikzpicture}}   \quad &+ \quad 
\raisebox{-0.44\height}{
\begin{tikzpicture}[scale=0.75*\picSc]
	\doublePlainLoop{2}{0}{2}
	\gluonLoop{2.7}{-0.1}{0.8}
	\gluonLine{0}{0}{1.9}{0}
	\blobNode{0}{0}{\bigBlobSc}{$E$}
	\blobNode{4}{0}{\bigBlobSc}{$E$}
\end{tikzpicture}}  \quad = \quad 3\left[d+1+\frac{2}{d-3}\right]\left[\mathcal{E}^0_1 (t)\right]^2\nonumber\\
&-6\left[1+\frac{2}{d-3}\right]\mathcal{E}^1_1(t) \mathcal{E}^1_2(t)+\Bigg\lbrace \left[(d-3)d-1-\frac{1}{d-3}\right]\left[\mathcal{E}^0_1 (t)\right]^2\nonumber \\
&-2\left[d-1-\frac{1}{d-3}\right]\mathcal{E}^1_1(t) \mathcal{E}^1_2(t) \Bigg  \rbrace(1-\xi)- \frac{d}{2}(d-2)\left[\mathcal{E}^0_1 (t)\right]^2 (1-\xi)^2 , \\
\raisebox{-0.44\height}{
\begin{tikzpicture}[scale=0.75*\picSc]
	\doublePlainLoop{2}{0}{2}
	\gluonLine{0}{0}{4}{0}
	\gluonArc{0}{2}{1.5}{-5}{-90}
	\blobNode{0}{0}{\bigBlobSc}{$E$}
	\blobNode{4}{0}{\bigBlobSc}{$E$} 
\end{tikzpicture}} \quad &+ \quad \raisebox{-0.44\height}{
\begin{tikzpicture}[scale=0.75*\picSc]
	\doublePlainLoop{2}{0}{2}
	\gluonLine{0}{0}{4}{0}
	\gluonArc{4}{2}{1.6}{185}{270}
	\blobNode{0}{0}{\bigBlobSc}{$E$}
	\blobNode{4}{0}{\bigBlobSc}{$E$} 
\end{tikzpicture}} \quad = \quad \left[\frac{4d}{d-2} + 2d(1-\xi)\right]\mathcal{E}^1_1(t) \mathcal{E}^1_2(t), \label{diagEW}\\
\raisebox{-0.44\height}{
\begin{tikzpicture}[scale=0.75*\picSc]
	\doublePlainLoop{2}{0}{2}
	\gluonArc{0}{2}{1.5}{-5}{-90}
	\gluonArc{4}{2}{1.5}{185}{270}
	\blobNode{0}{0}{\bigBlobSc}{$E$}
	\blobNode{4}{0}{\bigBlobSc}{$E$}
\end{tikzpicture}}  \quad &= \quad 0, \\
\raisebox{-0.44\height}{
\begin{tikzpicture}[scale=0.75*\picSc]
	\doublePlainLoop{2}{0}{2}
	\gluonArc{0}{3}{3}{-25}{-42}
	\gluonArc{0}{3}{3}{-52}{-90}
	\gluonArc{4}{3}{3}{205}{270}
	\blobNode{0}{0}{\bigBlobSc}{$E$}
	\blobNode{4}{0}{\bigBlobSc}{$E$}
\end{tikzpicture}} \quad  &= \quad 0, \\
\raisebox{-0.44\height}{
\begin{tikzpicture}[scale=0.75*\picSc]
	\doublePlainLoop{2}{0}{2}
	\gluonLine{0}{0}{4}{0}
	\gluonLine{0.5}{1}{3.5}{1}
	\blobNode{0}{0}{\bigBlobSc}{$E$}
	\blobNode{4}{0}{\bigBlobSc}{$E$} 
\end{tikzpicture}} \quad &= \quad (d-1)\left[\frac{2}{d-2}+1-\xi\right]\left[\mathcal{E}^0_{2}(t)-\mathcal{E}^0_{2}(0)\right]\mathcal{E}^{2}_{1}(t). \label{eq:wwdiag}
\end{align}

The lower and upper correlators give mostly identical contributions. 
As discussed in section \ref{nonfact}, there is only one topology, the one featuring a Wilson line connected to a three-point vertex, in which they differ, and the difference appears when considering the zero mode. 
This zero-mode contribution is also the source of the breaking of the $t \to 1/T-t$ symmetry in the lower and upper correlators.
Indeed, the zero-mode contribution to this diagram gives a result of the form of $t\mathcal{Z}^1_{011}(t)$ for the upper correlator
and $-(1/T-t)\mathcal{Z}^1_{011}(t)$ for the lower correlator. 
The two contributions are different, which is the origin of the asymmetry.
However, the  two contributions are related by the transformation $t \to 1/T-t$ (note that $\mathcal{Z}^1_{011}(1/T-t) = \mathcal{Z}^1_{011}(-t) = -\mathcal{Z}^1_{011}(t)$), which makes their sum symmetric.
Additionally, this is the only topology without a closed-form value. 
The contributions from this topology read for the upper and lower correlator, respectively,
\begin{align}
\raisebox{-0.44\height}{
\begin{tikzpicture}[scale=0.75*\picSc]
	\doublePlainLoop{2}{0}{2}
	\gluonLine{0}{0}{4}{0}
	\gluonLine{2}{1.9}{2}{0}
	\blobNode{0}{0}{\bigBlobSc}{$E$}
	\blobNode{4}{0}{\bigBlobSc}{$E$}
\end{tikzpicture}} &\quad  = \quad \left[d-1+\frac{2}{d-3}\right]\left[\mathcal{E}^0_1 (t)-\mathcal{E}^0_1 (0)\right]\mathcal{E}^0_1 (t) \nonumber\\ 
&-2\left[3+\frac{2}{d-2}+\frac{6}{d-3}\right]\mathcal{E}^1_1(t)\mathcal{E}^1_2(t)-4\left[1+\frac{2}{d-3}\right]\left[\mathcal{E}^0_2(t)-\mathcal{E}^0_2(0)\right]\mathcal{E}^{2}_{1}(t)  \nonumber \\
&-i\left[d-1+\frac{2}{d-3}\right]\int_{0}^{t}\mathrm{d}t'\,\mathcal{E}_{1}^{1}\left(t'\right)\mathcal{E}_{1}^{0}\left(t-t'\right)\nonumber \\ &-\left\lbrace 2d\mathcal{E}^1_1(t)\mathcal{E}^1_2(t) + 2(d-1)\left[\mathcal{E}^0_2(t)-\mathcal{E}^0_2(0)\right]\mathcal{E}^{2}_{1}(t)\right\rbrace(1-\xi), \label{eq:w3gewe}\\
\raisebox{-0.44\height}{
\begin{tikzpicture}[scale=0.75*\picSc]
	\doublePlainLoop{2}{0}{2}
	\gluonLine{0}{0}{4}{0}
	\gluonLine{2}{-1.9}{2}{0}
	\blobNode{0}{0}{\bigBlobSc}{$E$}
	\blobNode{4}{0}{\bigBlobSc}{$E$}
\end{tikzpicture}}  &\quad  = \quad \left[d-1+\frac{2}{d-3}\right]\left[\mathcal{E}^0_1 (t)-\mathcal{E}^0_1 (0)\right]\mathcal{E}^0_1 (t) \nonumber\\
&-2\left[3+\frac{2}{d-2}+\frac{6}{d-3}\right]\mathcal{E}^1_1(t)\mathcal{E}^1_2(t)-4\left[1+\frac{2}{d-3}\right]\left[\mathcal{E}^0_2(t)-\mathcal{E}^0_2(0)\right]\mathcal{E}^{2}_{1}(t) \nonumber \\
&-i\left[d-1+\frac{2}{d-3}\right]\int_{t}^{1/T-t}\!\!\!\mathrm{d}t'\,\mathcal{E}_{1}^{1}\left(t'\right)\mathcal{E}_{1}^{0}\left(t-t'\right)\nonumber \\
&-\left\lbrace 2d\mathcal{E}^1_1(t)\mathcal{E}^1_2(t) + 2(d-1)\left[\mathcal{E}^0_2(t)-\mathcal{E}^0_2(0)\right]\mathcal{E}^{2}_{1}(t)\right\rbrace(1-\xi).  \label{eq:w3geew} 
\end{align}

To obtain the symmetric correlator, some changes and additions are needed. 
The three contributions unique to the symmetric correlator are
\begin{align}
\raisebox{-0.44\height}{
\begin{tikzpicture}[scale=0.75*\picSc]
	\doublePlainLoop{2}{0}{2}
	\gluonArc{0}{2}{2}{-5}{-90}
	\gluonArc{4}{-2}{2}{90}{175}
	\blobNode{0}{0}{\bigBlobSc}{$E$}
	\blobNode{4}{0}{\bigBlobSc}{$E$},
\end{tikzpicture}}  \quad &+ \quad
\raisebox{-0.44\height}{
\begin{tikzpicture}[scale=0.75*\picSc]
	\doublePlainLoop{2}{0}{2}
	\gluonArc{0}{-2}{2}{0}{90}
	\gluonArc{4}{+2}{2}{180}{270}
	\blobNode{0}{0}{\bigBlobSc}{$E$}
	\blobNode{4}{0}{\bigBlobSc}{$E$},
\end{tikzpicture}} \quad = \quad 0, \\
\raisebox{-0.44\height}{
\begin{tikzpicture}[scale=0.75*\picSc]
	\doublePlainLoop{2}{0}{2}
	\gluonLine{0}{0}{4}{0}
	\gluonLine{2}{1.9}{2}{0.3}
	\gluonLine{2}{-0.3}{2}{-1.9}
	\blobNode{0}{0}{\bigBlobSc}{$E$}
	\blobNode{4}{0}{\bigBlobSc}{$E$} 
\end{tikzpicture}}  \quad &= \quad -(d-1)\left[\frac{2}{d-2}+1-\xi\right]\left[\mathcal{E}^0_2(t)-\mathcal{E}^0_2(0)\right]\mathcal{E}^{2}_{1}(t). \label{eq:wwewew}
\end{align}
The overall coupling and color factor is $g_s^4 d_A (N_c^2-4)$ ($g_s^4 d_A (N_c^2-4) N_f/N_c$ for the quark loop diagram) for the symmetric correlator with $x_{abc} = d_{abc}$,
and $g_s^4 d_A N_c^2$ ($g_s^4 d_A N_c N_f$ for the quark loop diagram) for the symmetric correlator with $x_{abc} = f_{abc}$.
There are four diagrams with the topology of \eqref{diagEW} contributing to the symmetric correlator, up from two for the upper or lower correlator.
Nevertheless, because they come with an extra factor $1/2$, their contribution is the same as for the upper or lower correlator apart from the overall color factor.
There are two diagrams with the topology of \eqref{eq:wwdiag} contributing to the symmetric correlator with no extra factors;  
the additional diagram exactly cancels the contribution from diagram \eqref{eq:wwewew}.
The diagrams \eqref{eq:w3gewe} and \eqref{eq:w3geew} both contribute to the symmetric correlator.
These diagrams also come with an extra factor $1/2$ for the case of the symmetric correlator.
The asymmetric parts embedded in the time convolution of the sum-integrals $\mathcal{E}^1_1$ and $\mathcal{E}^0_1$, see sections \ref{nonfact} and \ref{sec:zeromode},
sum up to give a contribution that is eventually symmetric under $t \to 1/T-t$.
We elaborate more on the symmetric correlator in section \ref{sec:resultsS}.

In all of the above diagrams, the vanishing results arise from symmetry considerations. 
In the case of vanishing mirrored pairs of diagrams, the diagrams  also vanish individually.

\subsection{Results for the upper and lower correlators}
\label{sec:resultsUL}
Adding the contributions of all diagrams, we obtain the nonrenormalised expression at NLO for the different correlators. 
For $I\in\lbrace U,L \rbrace$, the $d$-dimensional result reads
\begin{align}
      \langle EE \rangle_I &= - d_A g_s^2 T^{-4}(d-1) \mathcal{E}^2_1(t) + d_A N_c g_s^4T^{-4} \Bigg\lbrace 2\left(d+\frac{3}{d-3}\right)\left[\mathcal{E}^0_1 (t)\right]^2 \nonumber\\
      &-\frac{1}{2}\left[d\left(d^2-4d+9\right)-2+\frac{12}{d-3}\right]\mathcal{E}^0_1 (t)\mathcal{E}^0_1 (0)\nonumber\\
      &-4\left(\frac{d-3}{d-2}+\frac{4}{d-3}\right)\mathcal{E}^1_1 (t)\mathcal{E}^1_2 (t)-2\left(\frac{d-3}{d-2}+\frac{4}{d-3}\right)\left[\mathcal{E}^0_2 (t)-\mathcal{E}^0_2 (0)\right]\mathcal{E}^2_1 (t)\nonumber \\
      &-i\left(d-1+\frac{2}{d-3}\right)\int_{W_I}\!\!\mathrm{d}t'\,\mathcal{E}^{1}_{1}(t')\mathcal{E}^{0}_{1}(t-t')\Bigg\rbrace  \nonumber \\
      & + d_A N_f g_s^4 T^{-4}\Bigg\lbrace (d-1)\left[\mathcal{E}^0_1 \left(t\right)\right]^2+(d-1)4^{2-d}\left[\mathcal{E}^0_1 \left(\frac{t}{2}\right)\right]^2\nonumber \\
      &-(d-1)2^{3-d}\mathcal{E}^0_1 \left(\frac{t}{2}\right)\mathcal{E}^0_1 \left(t\right)+\left(2^{2-d}-1\right)\left[(d-3)d+4+\frac{4}{d-3}\right]\mathcal{E}^0_1 \left(t\right)\mathcal{E}^0_1 \left(0\right)\nonumber \\
      &-\left(1+\frac{2}{d-3}\right)\left[2\mathcal{E}^1_1(t)-2^{2-d}\mathcal{E}^1_1\left(\frac{t}{2}\right)\right]\left[\mathcal{E}^1_2(t)-2^{3-d}\mathcal{E}^1_2\left(\frac{t}{2}\right)\right]\Bigg\rbrace ,
      \label{eq:eweresult}
\end{align}
where $\displaystyle \int_{W_I}$ indicates integration over one of the Wilson lines: $\displaystyle \int_{W_U} = \int_0^t $ and $\displaystyle \int_{W_L} = \int_t^{1/T}$.
We see that the complete result is gauge independent, i.e. the dependence on the gauge parameter $\xi$ has cancelled, as expected, in the sum of all diagrams.

There is an ultraviolet divergence in the expression of the correlators when they are evaluated near $d=3$. 
Using eqs. \eqref{appB:EINT} and \eqref{app: IxDiv}, the $O(1/\varepsilon)$ divergence of any of the correlators is given by     
\begin{align}
  \langle EE \rangle_I\Big|_{\text{order\;} 1/\varepsilon}
  = -\frac{d_A  \pi^2 }{3\varepsilon}\frac{g_s^4}{(4\pi)^2} (2N_f-11N_c)\left[\mathrm{cos}(2\pi Tt)+2\right]\mathrm{csc}^4(\pi Tt).
\label{NLOEEdiv}     
\end{align}
From eqs. \eqref{eq:LOEE} and \eqref{NLOEEdiv}, we see that the renormalisation of the gauge coupling is enough to renormalise the whole correlator at one loop.
Renormalisation proceeds then by renormalisation of the coupling $g_s$, with the one-loop relation between the bare coupling, $g_s$, and the renormalised coupling in the $\overline{\text{MS}}$ scheme, $g_{s,\overline{\text{MS}}}$, being 
\begin{equation}
  g_s^2 = g_{s,\overline{\text{MS}}}^2\left(\bar{\Lambda}\right)\left[1+\frac{g_{s,\overline{\text{MS}}}^2\left(\bar{\Lambda}\right)}{3\varepsilon(4\pi)^2}\left(\frac{e^{\gamma_E}\bar{\Lambda}^2}{4\pi}\right)^{-\varepsilon}\left(2N_f-11N_c\right)\right],
\label{gsMSbar}
\end{equation}
where $\bar{\Lambda}$ is the renormalisation scale.
After renormalisation, the finite result for the correlator can be extracted straightforwardly from eq. \eqref{eq:eweresult} by expanding the functions $\mathcal{E}^a_m(t)$ according to eq. \eqref{appB:EINT},
the convolution integral \eqref{appB:Ixrsubtr} according to eqs. \eqref{appB:B4}, \eqref{app: IxDiv} and \eqref{app:Ixfinite}, and the zero-mode contribution according to eq. \eqref{eq:zresult}.
The final expression can be found in appendix \ref{app: Correxpanded}.

\subsection{Results for the symmetric correlator}
\label{sec:resultsS}
We next discuss the evaluation of the symmetric correlator. 
This involves summing eqs. \eqref{eq:selfE}--\eqref{eq:wwewew} according to the prescriptions given after eq. \eqref{eq:wwewew}.
At NLO, the symmetric correlator shares the same topologies of diagrams as the fundamental correlator \eqref{eq:EEF}. 
Therefore, it coincides with the fundamental correlator up to Casimir scaling, i.e. up to a factor $(N_c^2-4)/N_c$ for the symmetric correlator with $x_{abc} = d_{abc}$, 
and $N_c$ for the symmetric correlator with $x_{abc} = f_{abc}$.
Indeed, we have both compared the (non-reduced) integral forms to those obtained in \cite{Burnier:2010rp}, as well as computed the spectral function associated with the correlator by Fourier-transforming it and taking the imaginary part of its analytic continuation to real time. 
For the former, we found agreement with \cite{Burnier:2010rp}, and for the latter we found agreement up to a term proportional to $\pi^2$, a discrepancy that has already been noted and clarified recently in \cite{Scheihing-Hitschfeld:2023tuz,delaCruz:2024cix}. 
Following again the discussion after eq. \eqref{eq:wwewew}, up to a rescaling of the color factor, and up to $O(\varepsilon)$-terms,
the symmetric correlator differs from the upper or lower correlator only in the asymmetry embedded in the convolution of the sum-integrals $\mathcal{E}^1_1$ and $\mathcal{E}^0_1$, see sections \ref{nonfact} and \ref{sec:zeromode}, i.e. 
\begin{align}
C\langle EE \rangle_S - \langle EE \rangle_U &= -\frac{i}{2\varepsilon} d_A N_c g_s^4  T^{-4}\mathcal{Z}^{1}_{011}(t) + O(\varepsilon) \nonumber\\
&=  -\frac{3}{\left(2\pi\right)^3} \pi d_A N_c g_s^4\left[\zeta\left(4,Tt\right)-\zeta\left(4,1-Tt\right)\right]+O(\varepsilon),
\label{DiffEESEEU}\\
C\langle EE \rangle_S - \langle EE \rangle_L &= \frac{i}{2\varepsilon} d_A N_c g_s^4  T^{-4}\mathcal{Z}^{1}_{011}(t)  + O(\varepsilon) \nonumber\\
&=  \frac{3}{\left(2\pi\right)^3} \pi d_A N_c g_s^4\left[\zeta\left(4,Tt\right)-\zeta\left(4,1-Tt\right)\right]+O(\varepsilon),
\label{DiffEESEEL}
\end{align}
where $C$ is a color rescaling factor: for the symmetric correlator with $x_{abc} = d_{abc}$, $C = N_c/\left(N_c^2-4\right)$ and for the symmetric correlator with $x_{abc} = f_{abc}$, $C =1/N_c$.
The difference is finite because ${\mathcal{Z}^{1}_{011}(t)}$ is $O(\varepsilon)$, see eq. \eqref{eq:zresult}, which we have used for the last set of equalities. The result in eq. \eqref{DiffEESEEU} agrees with the difference between the correlators in eq.~\eqref{eq:EEF}  and eq.~\eqref{defEEU} first reported in \cite{Eller:2019spw,Scheihing-Hitschfeld:2023tuz}.

\subsection{Results for the upper and lower correlator asymmetry}
\label{sec:resultsasy}
Thermal correlators in Euclidean space are naïvely expected to respect the symmetry properties of their composite fields.
As such, one would expect the chromoelectric correlators $\left \langle EE\right \rangle_I$ to be symmetric over the thermal circle, $\left \langle EE\right \rangle_I(t)=\left \langle EE\right \rangle_I\left(1/T-t\right)$.
However, this statement is only true for correlators constructed fully out of local operators, and lattice calculations have confirmed that this is not the case aside from the explicitly symmetric correlator $\langle EE\rangle_S(t)$. 
Since $\left \langle EE\right \rangle_U(t)=\left \langle EE\right \rangle_L(1/T-t)$, an equivalent way of stating the lack of symmetry is that the upper and lower correlators do not coincide. 
This leads to a number of complications such as introducing terms that are even in $\omega$ to the associated spectral function.
For lattice simulations a more serious issue is that it requires the standard integral formula connecting a spectral function $\rho^S$ and a symmetric Euclidean correlator~$\langle\mathcal{O}\rangle_{S}$, 
\begin{equation}\label{eq:inversion}
  \langle\mathcal{O}\rangle_{S} = \int_{0}^{\infty} 
  \!\frac{\mathrm{d} \omega}{\pi} \, \rho^S(\omega) \frac{\cosh\left(\frac{\omega}{2T}-\omega t\right)}{\sinh\left({\frac{\omega}{2T}}\right)}, 
\end{equation}
to be generalised, which further complicates the already challenging task of determining the spectral function from lattice data (see also relevant discussion in \cite{Scheihing-Hitschfeld:2023tuz}). 
For such reasons, it is important to understand the sources and form of the asymmetry. 

The results given in the previous sections contain the explicit expression for the asymmetry of the upper and lower chromoelectric correlators. 
The asymmetry appears starting at NLO and originates at that order from the diagrams shown in eqs. \eqref{eq:w3gewe} and \eqref{eq:w3geew}. 
It is given by (recall that $\left \langle EE \right \rangle_U (1/T-t) = \left \langle EE \right \rangle_L (t)$)
\begin{align}
\langle EE \rangle _\mathrm{A} (t) &\equiv \frac{1}{2}\left[\left \langle EE \right \rangle_U (t) - \left \langle EE \right \rangle_U (1/T-t)\right]\nonumber\\  
&= \frac{d_A N_c g_s^4T^{-4}}{2i}  \left(d-1+\frac{2}{d-3}\right) \mathcal{Z}^{1}_{011}(t) \nonumber \\
&=  \frac{3}{\left(2\pi\right)^3} \pi d_A N_c g_s^4\left[\zeta\left(4,Tt\right)-\zeta\left(4,1-Tt\right)\right]+O(\varepsilon). \label{eq:eweasym}
\end{align}
Because the function $\mathcal{Z}^1_{011}(t)$ is fully antisymmetric under $t \to 1/T-t$,
so is $\langle EE \rangle _\mathrm{A} (t)$.
Moreover, the antisymmetric part of $\left \langle EE \right \rangle_U (t)$ is 
$\langle EE \rangle_A(t)$, whereas the antisymmetric part of $\left \langle EE \right \rangle_L (t)$ is $-\langle EE \rangle_A(t)$.

At NLO, we have seen that the asymmetry is associated with the zero modes of a Wilson line.
The reason is clear: zero modes measure the length of the Wilson line, which is a distinguishing feature between the upper and lower correlators.
For the same reason, zero modes are expected to be a major source of the asymmetry also at higher orders.
If, as the NLO result seems to suggest, the zero modes are the only source of the asymmetry at all orders remains to be formally proved.

\subsection{Hard thermal loop resummation}
\label{sec:HTL}
There are no infrared divergences present in eq. \eqref{eq:eweresult}, a fact that has been previously established in \cite{Burnier:2008ia,Burnier:2010rp, Binder:2021otw}. 
This means that the correlator at NLO is not sensitive to contributions from the soft scale, i.e. contributions induced by momenta or energies of the order of the Debye mass.
Nevertheless, soft contributions have been computed in the literature.
The leading-order soft contributions to the spectral function have been first derived in \cite{Burnier:2010rp}.
They come from the hard thermal loop resummation of the diagrams shown in \eqref{gluonselfenergy}.
Up to Casimir scaling, the leading-order soft contribution to the spectral function is identical for all three adjoint correlators and the fundamental correlator.
As has been pointed out in \cite{Burnier:2010rp}, soft contributions to the correlator are of order $g_s^5$.
Indeed, the hard thermal loop resummation of the diagrams shown in \eqref{gluonselfenergy} provides the complete order $g_s^5$ contributions to the correlator,
as hard thermal loop resummed diagrams of the type \eqref{top1}-\eqref{top4} are suppressed by at least $g_s^2$. 
Hence, the correlator to order $g_s^5$ reads
\begin{equation}\label{eq: soft}
  \langle EE\rangle_{I}\Big|_{\text{to order\;} g_s^5}
  = \langle EE\rangle_{I} +  6 N_c T^{-4} \int_{0}^{+\infty} 
  \!\frac{\mathrm{d} \omega}{\pi}\, \rho^{\mathrm{soft}}(\omega) \frac{\cosh\left(\frac{\omega}{2T}-\omega t\right)}{\sinh\left({\frac{\omega}{2T}}\right)} ,
\end{equation}
where $\langle EE\rangle_{I}$ is the correlator at NLO given in eq. \eqref{eq:eweresult} and computed in this work, and $\rho^{\mathrm{soft}}(\omega)$ is the soft  spectral density. 
The soft spectral density can be read off from \cite{Burnier:2010rp}: it is precisely the sum of eqs. (4.3) and (4.6) given there. For the symmetric correlator with the chromoelectric fields multiplied by $x_{abc}=d_{abc}\,\,\left(x_{abc}=f_{abc}\right)$ the colour factor of the soft contribution in eq. \eqref{eq: soft} is $N_c^2-4$ ($N_c^2$) instead of $N_c$.
The numerical impact of the soft contribution to the correlators turns out to be very small.

\section{Discussion and comparison with lattice QCD}
\label{sec:discussion}
In this section, we evaluate numerically the correlators based on the NLO expression in eq.~\eqref{eq:eweresult} supplemented by the $g_s^5$ corrections discussed in section~\ref{sec:HTL},
expanded around $d=3$ and renormalised in the $\overline{\text{MS}}$ scheme, eq. \eqref{DiffEESEEU} for the symmetric correlator, and eq. \eqref{eq:eweasym} for the asymmetry.
For the renormalised coupling in the $\overline{\text{MS}}$ scheme,  $g_{s,\overline{\text{MS}}}^2 (\bar{\Lambda})$, we use the one-loop running coupling, as this is consistent with our precision. 
We set the renormalisation scale, $\bar{\Lambda}$, to the one-loop optimised scale following \cite{Burnier:2010rp,Brambilla:2022xbd}, 
\begin{equation} \label{eq:pmsscale}
    \bar{\Lambda}_\mathrm{PMS} = 4\pi T \exp \left[-\gamma_E - \frac{N_c-8N_f \ln 2}{2\left(11N_c-2N_f\right)} \right]. 
\end{equation}
We estimate the accuracy and applicability of our results by varying the scale about this central value by a factor of two to obtain uncertainty bands. 
Even when we choose to show only this central value for visual clarity, it should be understood as a central value within a band. 
Our choice of scale is not unique, with non-optimised choices of scales leading to larger uncertainty bands at the center of the thermal circle.
At the edges of the thermal circle, scales proportional to $1/t$ or $1/(1/T-t)$ dominate over scales proportional to $T$ and may be more appropriate.
The breakdown of perturbation theory for the renormalization scale $\bar{\Lambda}_\mathrm{PMS}$ is signalled by the sum of the NLO symmetric and antisymmetric contributions becoming larger than the LO term at $T t  \lesssim 1$.\footnote{
At $Tt \gtrsim 0$ the two contributions have opposite signs and tend to cancel each other out.}
 Although the breakdown of perturbation theory at the edge of the thermal circle may be cured by choosing a renormalization scale proportional to  $1/(1/T-t)$,
we have made the choice to use the $t$-independent scale $\bar{\Lambda}_\mathrm{PMS}$ everywhere in order to facilitate comparisons with existing literature and lattice calculations. 
In what follows, we set $N_c=3$, and consider various values of the temperature and $N_f$. 

\begin{figure}[ht]
\centering
\includegraphics[width=0.9\textwidth]{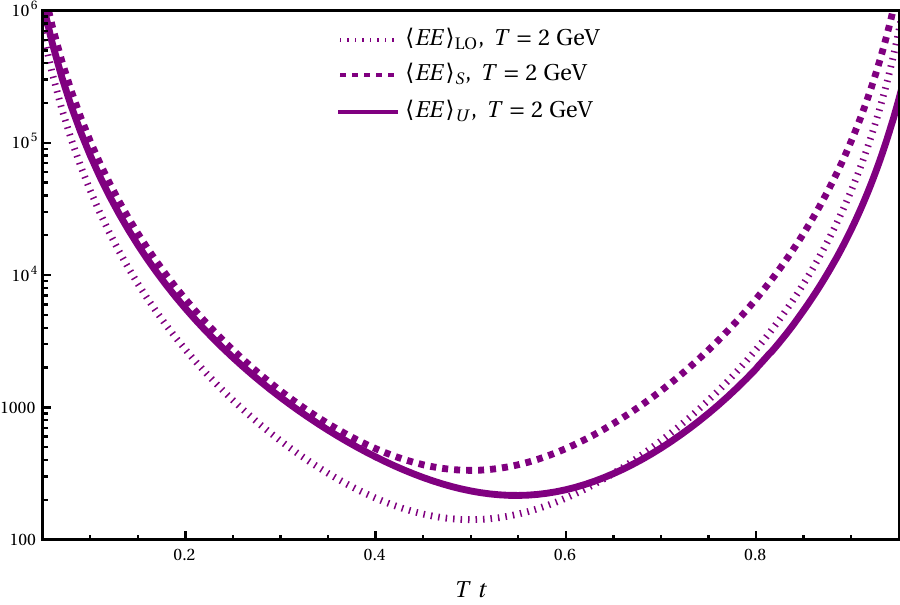}
\caption{The  upper chromoelectric correlator \eqref{defEEU} and the symmetric correlator with $x_{abc}=d_{abc}$ \eqref{defEES} as a function of $Tt$ evaluated at $T=2\, \mathrm{GeV}$, $N_c = N_f = 3$, displayed  on a logarithmic scale. 
The LO correlator, $\langle EE \rangle_{\mathrm{LO}}$,  is shown for comparison. 
The renormalisation scale is set to~$\bar{\Lambda}_{\mathrm{PMS}}$.  
\label{fig:ppmainewe1}}
\end{figure}

Figure \ref{fig:ppmainewe1} shows the upper and symmetric correlators at a temperature of $T=2\,\mathrm{GeV}$, with $N_f=3$. 
The symmetric correlator is shifted upwards compared to the LO one. 
To keep the results readable, we show here only the curves for the central value of the renormalisation scale.
While its numerical values are within the same range of the symmetric correlator, there is a qualitative difference in the upper correlator, with its minimum being clearly shifted towards the edge of the thermal circle. 

\begin{figure}[ht]
\centering
\includegraphics[width=0.95\textwidth]{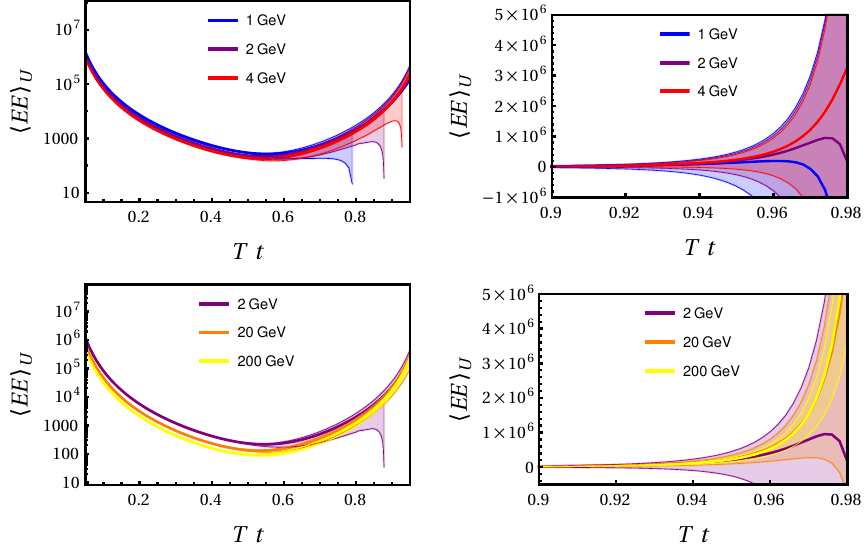}
\caption{The upper chromoelectric correlator \eqref{defEEU} as a function of $Tt$ evaluated at a range of temperatures for $N_c = N_f = 3$, with the renormalisation scale set at $\bar{\Lambda}_{\mathrm{PMS}}$. 
The scale is logarithmic for the entire thermal circle in the left panel and linear for the edge of the circle in the right panel, where the breakdown of perturbation theory can be seen as rapidly widening bands.}
\label{fig:ewetemps}
\end{figure}

In figure \ref{fig:ewetemps}, we elaborate on the effect of the asymmetry by showing the upper correlator, including uncertainty bands around the optimised scale, at a variety of temperatures, again with $N_f=3$. 
We can see that for our time-independent choice of the renormalisation scale, the part of the thermal circle where the perturbative result can be trusted extends further as the temperature is increased.
Good convergence for large values of $Tt$ requires extremely high temperatures, but for lower values of $Tt$ the results are stable also at relatively low temperatures.

\begin{figure}[ht]
\centering
\includegraphics[width=0.9\textwidth]{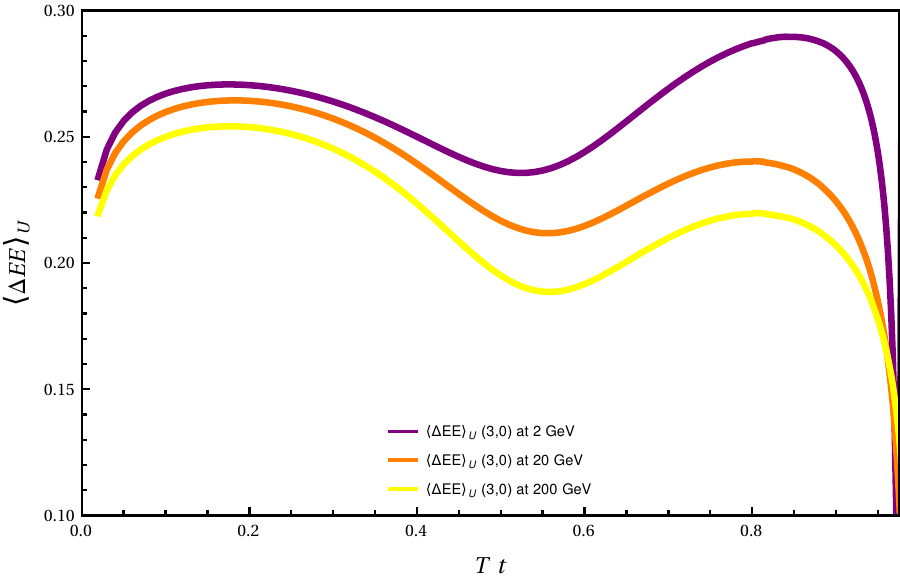}
\caption{The relative difference $\langle \Delta EE \rangle _U (N_f,N_f') \equiv | \langle EE \rangle_U(N_f)-\langle EE \rangle_U(N_f')|/\langle EE \rangle_U(N_f)$ of two upper correlators
  with a different number of quark flavours, $N_f=3$ and $N_f'=0$, as a function of $Tt$ evaluated for a range of temperatures for $N_c=3$. 
  The renormalisation scale is set to $\bar\Lambda_{\mathrm{PMS}}$.
  \label{fig:nffig}}
\end{figure}

As current state-of-the-art lattice calculations are still performed in pure Yang--Mills theory,
we also consider the effect of varying the number of fermions from the value $N_f=3$, most pertinent to the QGP produced in colliders, to $N_f=0$. 
The general effect of an increasing number of fermions is a slight upwards shift of the correlators when evaluated at the central renormalisation scale defined by eq. \eqref{eq:pmsscale}. 
Note that, as the renormalisation scale itself depends on $N_f$, a change in $N_f$ affects the coupling and therefore all contributions, not only the quark loop diagram. 
This causes an increase in the renormalisation-scale uncertainty as $N_f$ increases, particularly towards the edges of the thermal circle. 
This increase is not qualitatively significant, and happens due to our choice of the renormalisation scale. 
While the correlators do not change qualitatively, there is some quantitative change: 
for $N_f=3$ there is a change of roughly $25$--$30\%$ in the correlator at NLO with respect to the pure gauge case at the considered temperatures. 
As such, we are led to believe that for future QCD precision computations on the lattice, fermions ought to be included.
We show the associated relative difference between the $N_f=3$ and pure gauge cases for the upper correlator in figure \ref{fig:nffig}. 
The difference is uniform on most of the thermal circle, only dipping towards zero at the very end.
We observe the same general behaviour for a range of~$N_f$. 

\begin{figure}[ht]
\centering
\includegraphics[width=0.9\textwidth]{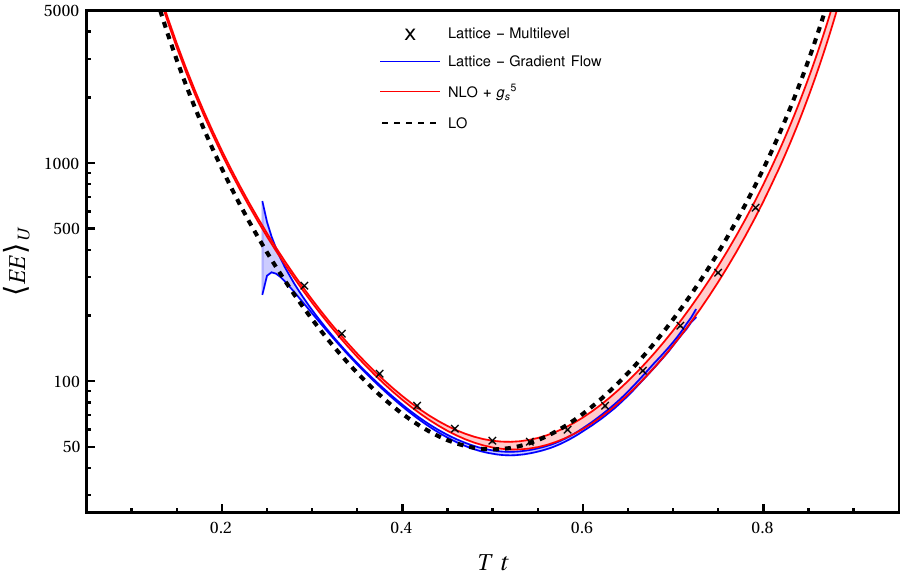}
\caption{Comparison between the lattice data obtained using gradient flow (blue band) and multilevel algorithm (points) of \cite{Brambilla:2025cqy} and the
  $g_s^5$ computation from  this work (red band) for the upper correlator $\langle EE \rangle _U (t)$ at $N_c=3$, $N_f=0$ and $T = 10^4 T_c \approx 3 \times 10^3 \mathrm{GeV}$.
  The perturbative error band is obtained by varying the renormalisation scale by a factor of two around the central value $\bar{\Lambda}_{\mathrm{PMS}}$.
\label{fig:compplot}}
\end{figure}

Recent results on the lattice \cite{Brambilla:2025cqy} using the gradient flow method \cite{Narayanan:2006rf,Luscher:2009eq,Luscher:2010iy} 
and multilevel algorithm \cite{Luscher:2001up} 
suggest that the correlator responsible for the singlet-octet transition is asymmetric on the thermal circle. 
In figure \ref{fig:compplot}, we compare our perturbative correlator \eqref{eq:eweresult} supplemented with the $g_s^5$ contributions discussed in section \ref{sec:HTL}
with the lattice results \cite{Brambilla:2025cqy} (interpolated between measurement points in the gradient flow case) on a logarithmic scale. 
The comparison is performed in pure gauge theory, i.e. $N_f$ = 0, and at sufficiently high temperatures for perturbation theory to be reliable. 
Specifically, we have used temperatures of approximately $10^4T_c$, where $T_c \sim 300\, \mathrm{MeV}$ for $N_f=0$. 
The lattice results have the same functional shape as the NLO correlator, and in particular they are both skewed towards the end of the thermal circle. 
In the perturbative calculation, the values for large $Tt$ exhibit increasing errors due to our choice of the renormalisation scale.
For the lattice results with the gradient flow method, the errors, which include statistical and systematic errors, increase instead for small~$Tt$. 
This is due to the fact that, as the value of $Tt$ becomes smaller, the statistical error increases significantly, because of limitations in extracting the short-distance (small-$Tt$) behaviour when applying the gradient flow method. The errors of the lattice results with the multilevel algorithm are mostly statistical. 
Accounting for the errors, the perturbative curve agrees well with the lattice data.

In addition to the perturbative errors, another source of uncertainty in our weak-coupling computation is the numerical evaluation of the convolution integral (see appendix~\ref{app:ixr}).
In practice, the numerical method required is simple enough far from the edges of the thermal circle, but the cancellations of divergences become very delicate towards the edges.
Problems related to numerics are far simpler to identify than the perturbative uncertainty,
as the resulting numerical instabilities lead to e.g. the zero-mode-subtracted integral (and the symmetric correlator) not being exactly symmetric despite having been  explicitly defined to exclude the zero-modes. 
For this and the other reasons discussed in the previous paragraphs, we do not show contributions that are less than $0.05$ away from the edges of the thermal circle (i.e.  for $Tt<0.05$ and $Tt>0.95$).

\begin{figure}[ht]
\centering
\includegraphics[width=0.95\textwidth]{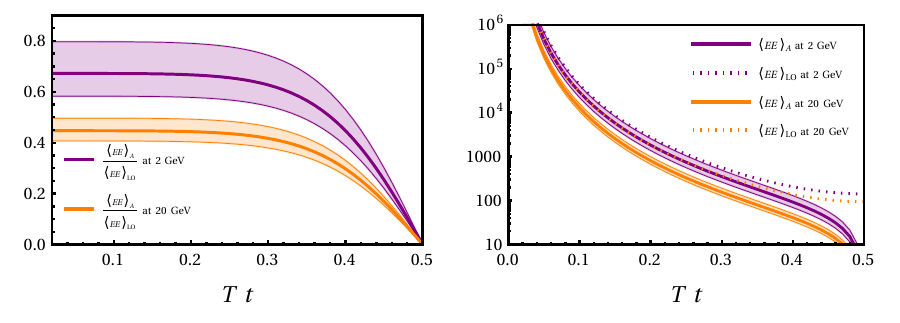}
\caption{The asymmetry \eqref{eq:eweasym} as a function of $Tt$ evaluated for $N_c = N_f = 3$ at $T=2\, \mathrm{GeV}$ (purple band) and $T=20\, \mathrm{GeV}$ (orange band). 
  The LO contributions are used as normalisations in the left plot and shown as dashed lines (no scale variation included) in the right plot.
  The widths of the bands are determined by renormalisation scale variations around the central value $\bar\Lambda_{\mathrm{PMS}}$ represented by the thick lines.
  \label{fig:asymfig}}
\end{figure}

\begin{figure}[ht]
\centering
\includegraphics[width=0.9\textwidth]{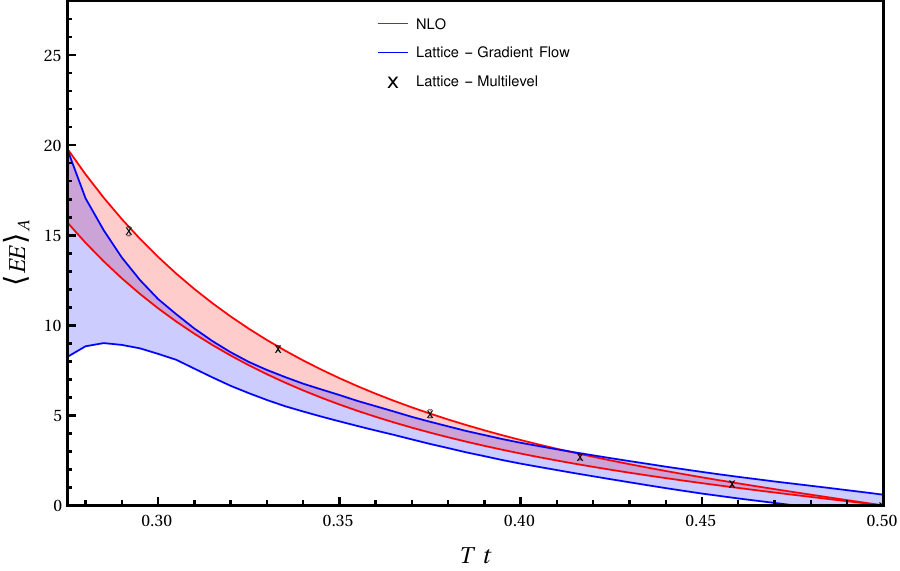}
\caption{Comparison between the asymmetry \eqref{eq:eweasym} (red band) and the lattice data of \cite{Brambilla:2025cqy} as in figure \ref{fig:compplot}. 
  The perturbative and lattice curves are both evaluated as functions of $Tt$ for $N_f =0$ at $T = 10^4 T_c \approx 3 \times 10^3 \mathrm{GeV}$.
  \label{fig:asymfig2}}
\end{figure}

The leading result for the asymmetry of the chromoelectric correlator \eqref{eq:eweasym} with $N_f=3$ at $T=2\, \mathrm{GeV}$ can be seen in the right panel of figure \ref{fig:asymfig} on a logarithmic scale. 
Since $\langle EE \rangle_A(t)$ is fully antisymmetric, we only show the first half of the thermal circle where the antisymmetric part has the same sign as the LO term.
The result is numerically significant despite being suppressed by a power of $g_s^2$. 
We note that, while the behaviour does not change qualitatively with increased temperature, convergence does significantly improve. 
In the left panel, we compare the asymmetry of the NLO correlator to the complete LO term.
We see that, while still sizeable at $T=20\,\mathrm{GeV}$ (at most 50\% of the LO), the asymmetry is better behaved than at $T=2\,\mathrm{GeV}$ (at most 80\% of the LO). 
At low temperatures and far from the center of the thermal circle, not only is the antisymmetric part comparable to the LO contribution, but its renormalisation scale variation is also large. 
In figure \ref{fig:asymfig2}, we compare the perturbative prediction of eq. \eqref{eq:eweasym} with the lattice results of \cite{Brambilla:2025cqy} obtained in the pure gauge theory at $T = 10^4 T_c$.
The asymmetry of the NLO correlator appears to agree well with the nonperturbative lattice determination. 
It should be noted that at $Tt = 0.5$, eq. \eqref{eq:eweasym} vanishes.

\section{Conclusions}
Motivated by quarkonium dynamics, we set out to consider correlators of two chromoelectric fields joined by adjoint Wilson lines at finite temperature $T$ in Euclidean space,
separated by an imaginary-time distance $t$ in a general covariant gauge. 
We identify three distinct gauge-invariant correlators and use integration-by-parts techniques to compute these correlators to NLO, see section~\ref{sec:HTL}.
Our final results also incorporate soft effects due to the Debye mass scale and are accurate up to order $g_s^5$, see section \ref{sec:HTL}.
One correlator, which is symmetric under $t \to 1/T-t$, is equivalent to a previously studied correlator in the fundamental representation up to Casimir scaling.
For the other two, we have identified a specific zero-mode contribution responsible for the breaking of the $t \to 1/T-t$ symmetry.

At NLO, the antisymmetric contribution responsible of the asymmetry originates
exclusively from zero-modes of Wilson lines. 
Such an identification may help in the future to push the computation of the asymmetry to higher orders.
We compare our results with those of pure gauge lattice computations at extremely high temperatures, and observe good agreement, see figure \ref{fig:compplot}.
In particular, the perturbative asymmetry agrees with the asymmetric behaviour observed on the lattice, see figure \ref{fig:asymfig2}.  
Given that our results for varying the number of massless quark flavours in the medium indicate that the fermionic contribution is numerically significant, we argue that quarks should be included in future precision lattice calculations.
Finally, an obvious extension of the present work is the computation of the transport coefficients associated with the considered correlators.

\section*{Acknowledgements}
S.S. wishes to thank J. Österman for pointing out \cite{Davydychev:2022dcw} as a convenient reference for reducing two-loop integrals,
as well as one of the authors of \cite{Davydychev:2023jto}, P. Navarrete, for elaborating some details of the limitations of the results therein. 
We thank  S. Datta, M. Janer, J. Meyer-Steudte and P. Petreczky for fruitful discussions, particularly regarding the lattice-field-theoretical evaluation of the chromoelectric correlator. 
S.S. thanks P. Petreczky and the Brookhaven National Laboratory for hospitality during a visit when part of this research was conducted. The authors acknowledge the support of the DFG cluster of excellence ORIGINS funded by the DFG under Germany’s Excellence Strategy - EXC-2094-390783311, the STRONG-2020, as well as the European Union’s Horizon 2020 research and innovation program under grant agreement No. 824093. 
The work of N.B. is supported by the DFG Grant No. BR 4058/5-1 “Open Quantum Systems and Effective Field Theories for hard probes of hot and/or dense medium”.
N.B. acknowledges the European Research Council advanced grant  ERC-2023-ADG-Project EFT-XYZ.
S.S. is currently affiliated with the Institute of Space Sciences (ICE--CSIC) and Institut d’Estudis Espacials de Catalunya (IEEC),
and supported by Ministerio de Ciencia, Investigacion y Universidades (Spain) MCIN/AEI/10.13039/501100011033/ FEDER, UE, under the project PID2022-139427NB-I00,
and by the Spanish program Unidad de Excelencia Maria de Maeztu CEX2020-001058-M, but the bulk of this research was carried out while her affiliation was with TU Munich.

\clearpage
\appendix

\section{Conventions and Feynman rules}\label{app:conventions}
Below we set the conventions and Feynman rules that we use throughout this work. 
We work in Euclidean space in the $R_\xi$-family of gauges. 
We also use dimensional regularisation in $d=3-2\varepsilon$ dimensions and eventually renormalise in the $\overline{\text{MS}}$-scheme. 
For example, a rotationally symmetric bosonic one-loop sum-integral is written explicitly as
\begin{equation}
    \sumint_P f(P) \equiv T \sum_{p_0} \int_{\mathbf{p}} f(p_0,p) \equiv T\lambda(S^{d-1})\left( \frac{e^{\gamma_\mathrm{E}}\bar{\Lambda}^2}{4\pi} \right)^{\frac{3-d}{2}} \int_0^{\infty} \mathrm{d} p \, p^{d-1} \sum_{n\in\mathbb{Z}} f(2\pi Tn,p),
\end{equation}
where $P=\left(p_0,\mathbf{p}\right)$, $|\mathbf{p}|\equiv p$, $p_0\in \left\lbrace 2 \pi T n | n\in\mathbb{Z}\right\rbrace$ are the Matsubara modes,
$\displaystyle \int_{\mathbf{p}} \equiv \left(\frac{e^{\gamma_\mathrm{E}}\bar{\Lambda}^2}{4\pi} \right)^{\frac{3-d}{2}} $ 
$\displaystyle \times \int \frac{\mathrm{d}^dp}{(2\pi)^{d}}$,\footnote{
  The factor $\left[ {e^{\gamma_\mathrm{E}}\bar{\Lambda}^2}/{(4\pi)} \right]^{\varepsilon}$ prepares for the $\overline{\mathrm{MS}}$ renormalisation scheme and it cancels against
  $\left[ {e^{\gamma_\mathrm{E}}\bar{\Lambda}^2}/{(4\pi)} \right]^{-\varepsilon}$ accompanying the coupling, see \eqref{gsMSbar}.
}
$\bar{\Lambda}$ is the renormalisation scale,
$\lambda(S^{d-1})\equiv\left[\Gamma\left(d/2\right)\pi^{d/2}2^{d-1}\right]^{-1}$ is a normalised measure of the $(d-1)$-sphere,
and $\gamma_\mathrm{E}=0.5772\ldots $ is the Euler--Mascheroni constant.
For fermionic Matsubara sums (where  $p_0\in \left\lbrace \pi T (2n+1) | n\in\mathbb{Z}\right\rbrace$) 
\begin{equation}
    \sumint_{\{P\}} f(P) \equiv  T\lambda(S^{d-1})\left( \frac{e^{\gamma_\mathrm{E}}\bar{\Lambda}^2}{4\pi} \right)^{\frac{3-d}{2}} \int_0^{\infty} \mathrm{d} p \, p^{d-1} \sum_{n\in\mathbb{Z}} f\left[\pi T(2n+1),p\right],
\end{equation}
but in practice all fermionic sums are computed using the standard formula \eqref{eq:1lfermion}. 
The group-theoretical factors for the gauge group $\mathrm{SU}(N_c)$ are
\begin{align}
    &\delta^{aa} \equiv d_A = N^2_c-1,\qquad f^{abc}f^{abc} = N_c d_A,
\end{align}
where $N_c$ is the number of colours and  $f^{abc}$ are the $\mathrm{SU}(N_c)$ structure constants.

We list below the bare Euclidean Feynman rules for QCD that we use. 
The gluon (in $R_\xi$ gauge), the massless quark, and the ghost propagators are respectively
\begin{align}
\raisebox{0\height}{
\begin{tikzpicture}[scale=1.25*\picSc]         \gluonLine{0}{0}{4}{0} 
\end{tikzpicture}} &\quad = \quad \frac{\delta^{ab}}{P^2} \left[ \delta_{\mu\nu}-(1-\xi)\frac{P_\mu P_\nu}{P^2}\right] \equiv \delta^{ab} G_{\mu\nu}(P),\\
\raisebox{0\height}{
\begin{tikzpicture}[scale=1.25*\picSc]
    \quarkLine{0}{0}{4}{0}{0.5}
\end{tikzpicture}} &\quad = \quad \frac{-\slashed{P}}{P^2},\\
\raisebox{0\height}{
\begin{tikzpicture}[scale=1.25*\picSc]
    \ghostLine{0}{0}{4}{0}{0.5}
\end{tikzpicture}} &\quad = \quad \frac{\delta^{ab}}{P^2}.
\end{align}
The vertices read (in the three-gluon vertex all momenta are incoming)
\begin{align}
\raisebox{-0.44\height}{
\begin{tikzpicture}[scale=1.25*\picSc]
    \gluonLine{0}{0}{2}{0};
    \gluonLine{2}{0}{3}{2};
    \gluonLine{2}{0}{3}{-2};
    \node at (-0.3,0.5){$P$};
    \node at (3,2.5){$Q$};
    \node at(3,-2.5){$R$};
\end{tikzpicture}} &\quad = \quad ig_sf^{abc}\left[\delta_{\mu \nu}\left(P-Q\right)_\rho + \delta_{\nu \rho}\left( Q-R \right)_\mu+\delta_{\rho \mu}\left( R-P \right)_\nu \right],\\ \nonumber\\
\raisebox{-0.44\height}{
\begin{tikzpicture}[scale=1*\picSc]            \gluonLine{0}{2}{2}{0}
    \gluonLine{0}{-2}{2}{0}
    \gluonLine{2}{0}{4}{2}
    \gluonLine{2}{0}{4}{-2}
    \node at (0,2.5) {$P$};
    \node at (4,2.5) {$Q$};
    \node at (4,-2.5){$R$};
    \node at (0,-2.5) {$S$};
\end{tikzpicture}} &\quad = \quad -g_s^2\Bigg \lbrace f^{eab}f^{ecd} \left( \delta_{\mu\rho}\delta_{\nu\sigma}-\delta_{\mu\sigma}\delta_{\nu\rho} \right)\nonumber\\
&+f^{eac}f^{edb}\left(\delta_{\mu\sigma}\delta_{\nu\rho}-\delta_{\mu\nu}\delta_{\sigma\rho} \right)
+ f^{ead}f^{ebc}\left(\delta_{\mu\nu}\delta_{\sigma\rho}-\delta_{\mu\rho}\delta_{\sigma\nu} \right) \Bigg \rbrace, \\
\raisebox{-0.44\height}{
\begin{tikzpicture}[scale=1.25*\picSc]
    \gluonLine{0}{0}{2}{0}
    \quarkLine{2}{0}{3}{2}{0.5}
    \quarkLine{3}{-2}{2}{0}{-0.5}
    \node at (-0.3,0.5) {$P$};
    \node at (3,2.5) {$Q$};
    \node at (3,-2.5) {$R$};
\end{tikzpicture}} &\quad= \quad g_s\gamma_{\mu}T^a, \\
\raisebox{-0.44\height}{
\begin{tikzpicture}[scale=1.25*\picSc]
    \gluonLine{0}{0}{2}{0}
    \ghostLine{2}{0}{3}{2}{0.5}
    \ghostLine{3}{-2}{2}{0}{-0.5}
    \node at (-0.3,0.5) {$P$};
    \node at (3,2.5){$Q$};
    \node at(3,-2.5){$R$};
\end{tikzpicture}} &\quad = \quad  ig_sf^{abc}Q_{\mu},
\end{align}
where $g_s$ is the bare strong coupling.

Besides the use of standard Feynman rules, our calculations involve Wilson lines and chromoelectric field insertions. 
We define the chromoelectric field and the Wilson line in the adjoint representation as 
($D_\mu = \partial_\mu + ig_s A_\mu$)
\begin{align}
    &E^a_i \equiv F^a_{0i} = \partial_0A^a_i-\partial_iA^a_0-g_sf^{abc}A_{0}^bA_{i}^c,\\
    &U^{ab}(t,t') \equiv \left[\mathrm{P} \exp\left( -i g_s  \int_t^{t'} \mathrm{d}t''
    A_0(t'')  \right)\right]^{ab},
\end{align}
where $(A_\mu)_{ab}=(A_\mu^cT_c)_{ab}$ is the gauge field in the adjoint representation of $\mathrm{SU}(N_c)$ with $\left(T_c\right)_{ab}=-if_{abc}$, and P stands for path ordering.

\section{Evaluating two-loop sum-integrals} \label{app:ixr}
Most of the two-loop sum-integrals that we encounter can be factorised into one-loop integrals. 
We evaluate here the general finite one-loop integral using dimensional regularisation:
\begin{align}
\mathcal{E}^{a}_{m}(t) \equiv& \sumint_P e^{ip_0 t} \frac{p_0^{a}}{P^{2m}}=T\sum_{p_0}e^{ip_0 t} p_0^{a}I_{m}\left(p_0\right)\nonumber\\
=& T\left( \frac{e^{\gamma_\mathrm{E}}\bar{\Lambda}^2}{4\pi} \right)^{\frac{3-d}{2}}\frac{\Gamma\left(m-d/2\right)}{\left(4\pi \right)^{d/2}\Gamma\left( m\right)}\sum_{p_0}e^{ip_0 t} p_0^{a}\left(p_{0}^2\right)^{d/2-m}\nonumber \\
=&\frac{T^{4-2m+a}}{\left(2\pi\right)^{2m-a}}\frac{\Gamma\left(m-\frac{d}{2}\right)}{\Gamma\left(m\right)}\pi^{\frac{3}{2}}\left(\frac{e^{\gamma_{E}}\bar{\Lambda}^{2}}{4\pi^{2}T^{2}}\right)^{\frac{3-d}{2}}
\nonumber\\
&\qquad\times
\left[\mathrm{Li}_{2m-a-d}\left(e^{2\pi iTt}\right)+(-1)^{a}\mathrm{Li}_{2m-a-d}\left(e^{-2\pi iTt}\right)\right] \nonumber\\
&\!\!\!\!\!\!\!\!\!\!\!\!\!
\overset{d \to 3-2\varepsilon}{=}2^{a-2m}\pi^{a-2m+\frac{3}{2}}\frac{\Gamma\left(m-\frac{3}{2}\right)}{\Gamma(m)}T^{a-2m+4}
\nonumber\\
&\qquad\times
\left[\mathrm{Li}_{2m-a-3}\left(e^{2\pi iTt}\right)+(-1)^{a}\mathrm{Li}_{2m-a-3}\left(e^{-2\pi iTt}\right)\right]\nonumber\\
&+\nonumber 2^{a-2m}\pi^{a-2m+\frac{3}{2}}\frac{\Gamma\left(m-\frac{3}{2}\right)}{\Gamma(m)}T^{a-2m+4}
\nonumber\\
&\qquad\times
\bigg\lbrace2\left[\mathrm{Li}^{(1)}_{2m-a-3}\left(e^{2\pi iTt}\right)+(-1)^{a}\mathrm{Li}^{(1)}_{2m-a-3}\left(e^{-2\pi iTt}\right)\right]\nonumber\\
&\!\!\!\!\!\!\!\!\!\!+\left[\mathrm{H}_{m-\frac{5}{2}}+2\ln \left(\frac{\bar \Lambda }{4\pi T}\right)\right]\left[\mathrm{Li}_{2m-a-3}\left(e^{2\pi iTt}\right)+(-1)^{a}\mathrm{Li}_{2m-a-3}\left(e^{-2\pi iTt}\right)\right]\bigg\rbrace\varepsilon +O(\varepsilon^2),\label{appB:EINT} 
\end{align}
where we have used the standard textbook integral
\begin{equation}
  I_{n}\left(M\right)\equiv\int_{\boldsymbol{p}}\frac{1}{\left(\boldsymbol{p}^{2}+M^{2}\right)^n}=
  \left( \frac{e^{\gamma_\mathrm{E}}\bar{\Lambda}^2}{4\pi} \right)^{\frac{3-d}{2}}\frac{\left(M^2\right)^{d/2-n}\Gamma\left(n-d/2\right)}{\left(4\pi \right)^{d/2}\Gamma\left( n\right)},
\end{equation}
and the series definition of polylogarithm: $\displaystyle \mathrm{Li}_s(z) = \sum_{k=1}^{\infty}  k^{-s}z^k$. 
Moreover, we have defined $\displaystyle \mathrm{Li}^{(n)}_{s}\left(z\right) \equiv \frac{\partial^{n}}{\partial s^n}\mathrm{Li}_{s}\left(z\right)$, and $\mathrm{H}_n$ is the $n$th Harmonic number. 
In eq. \eqref{appB:EINT}, we display the $\varepsilon$ expansion for finite $\mathcal{E}^{a}_{m}(t)$, since most of the one-loop sum-integrals that we encounter are finite. 
The only exception is,
\begin{align}
    \mathcal{E}^{0}_{2}(0)\overset{d \to 3-2\varepsilon}{=}\frac{1}{16\pi^2\varepsilon}+ \frac{1}{8\pi^2}\left[\ln{\left(\frac{\overline{\Lambda}}{4\pi T}\right)}-\gamma_{E}\right]+\frac{\varepsilon}{8\pi^2}\left[\ln^2{\left(\frac{\overline{\Lambda}}{4\pi T}\right)}-\gamma_{1}+\frac{\pi^2}{8}\right]+O(\varepsilon^2),
\label{appB:E20INT}    
\end{align}
where $\gamma_E$ is the Euler constant and $\gamma_1$ is a Stieltjes constant that appears when expanding the Riemann zeta function.

By employing integration-by-parts identities, we are able to compute most of the two-loop integrals appearing in the NLO correlators in closed form as products of two one-loop integrals of the type ${\cal E}^a_m$. 
We note that  ${\cal E}^a_m$ has vanishing zero modes in dimensional regularisation. 
We now focus on the one exception to this rule: The convolution integral defined in the second line of eq. \eqref{eq:Ixr}. 
To better isolate the divergence and have common results for all correlators, we choose to subtract the zero-mode contribution from \eqref{eq:Ixr} and define,
\begin{align} \label{appB:Ixrsubtr}
I(t) &\equiv \frac{i}{2}\int_{0}^{t}\mathrm{d}t'\,\mathcal{E}_{1}^{1}\left(t'\right)\mathcal{E}_{1}^{0}\left(t-t'\right)-\frac{t}{2} iT\mathcal{Z}^{1}_{011}(t). 
\end{align}
$I(t)$ is symmetric under $t \to T -1/t$.
To evaluate $I(t)$ numerically, we must remove the ultraviolet divergence present in the integral. 
The functions $\mathcal{E}^1_1(t')$ and $\mathcal{E}^0_1(t-t')$ are finite in $d=3$, but their convolution is divergent near the endpoints of the integration domain $(0,t)$. 
In order to compute the divergence, we perform a multivariate Taylor expansion of the integrand, treating the two convoluted sum-integrals as independent
and removing (and adding) sufficiently many terms to render the difference finite everywhere on the domain. 
The added counterterms are no longer coupled, and can be integrated and subsequently evaluated in a closed form. Explicitly, we get
\begin{align}
I(t) &= \frac{i}{2}\int_0^t \mathrm{d}t' \bigg [ \mathcal{E}^1_1 (t') \mathcal{E}^0_1 (t-t') - \mathcal{E}^1_1 (t) \mathcal{E}^0_1 (t-t') - \mathcal{E}^0_1 (t) \mathcal{E}^1_1 (t') \nonumber\\
&+i(t-t')  \mathcal{E}^2_1 (t)\mathcal{E}^0_1 (t-t') + it'  \mathcal{E}^1_1 (t) \mathcal{E}^1_1 (t')+\frac{\left(t'\right)^2}{2}   \mathcal{E}^2_1 (t)\mathcal{E}^1_1 (t') \bigg ] \nonumber\\
& +\frac{1}{2} \left\lbrace -it T\mathcal{Z}^{1}_{011}(t)  -it \mathcal{E}^1_1 (t) \mathcal{E}^0_1 (t) -\frac{t^2}{2}\mathcal{E}^2_1 (t) \mathcal{E}^0_1 (t) \right\rbrace \nonumber\\
& -\frac{2}{d-2} \left\lbrace \mathcal{E}^1_1(t) \left[\mathcal{E}^1_2(t) -\mathcal{E}^1_2(0)\right]+\mathcal{E}^2_1(t) \left[\mathcal{E}^0_2(t) -\mathcal{E}^0_2(0)\right] -it \mathcal{E}^2_1(t)\mathcal{E}^1_2(t)  \right\rbrace.
\end{align}
The expression under the integration sign is now finite, as can be readily confirmed numerically, whereas the last two lines contain only known functions that can be evaluated in closed form with the formulas given above.

To obtain a useful result in the calculation at hand, 
$I(t)$ must be evaluated to $O(\varepsilon)$, as it enters the correlator multiplied by $1/(d-3)$ (see  eq. \eqref{eq:eweresult}). 
To do so, we expand $I(t)$ in powers of $\varepsilon$:
\begin{equation}
  \frac{I(t)}{T^4} = \frac{1}{\varepsilon} \left[I\right]^{(-1)}(t)+\left[I\right]^{(0)}(t)+\varepsilon\left( \left[I\right]^{(1)}(t)
  + \left[I\right]^{(1,\bar\Lambda)}(t) \ln \frac{\bar \Lambda}{T}\right) + O(\varepsilon^2). 
\label{appB:B4}
\end{equation}
The first two terms can be evaluated in closed form. 
This is to be expected as they are required to cancel the ultraviolet divergence, which is known in closed form. 
We find
\begin{align}\label{app: IxDiv}
\left[I\right]^{(-1)}(t) \,\,&= -\frac{\csc^4\left( \pi Tt\right)}{16} \left[\cos(2\pi Tt\right) + 2], \\
\left[I\right]^{(0)}(t) \,\,\,\,\,&=-\frac{\csc^4\left( \pi Tt\right)}{16}\Bigg \lbrace \frac{3}{2}+2\left[\gamma_E +2\ln \left(\frac{\bar \Lambda }{2T}\right)
  + \ln\left(1-e^{2\pi i Tt}\right)+\ln\left(1-e^{-2\pi i Tt}\right)\right] \nonumber\\
&\times \left[\cos\left(2\pi Tt\right) + 2 \right]\Bigg \rbrace -\frac{1}{2} \left[ \mathrm{Li}_
{-3}^{(1)} \left(e^{2\pi i Tt}\right)+\mathrm{Li}_{-3}^{(1)} \left(e^{-2\pi i Tt}\right)\right].
\label{app:Ixfinite} 
\end{align}
We have confirmed these expressions numerically.

\begin{figure}
\centering
\includegraphics[width=0.9\textwidth]{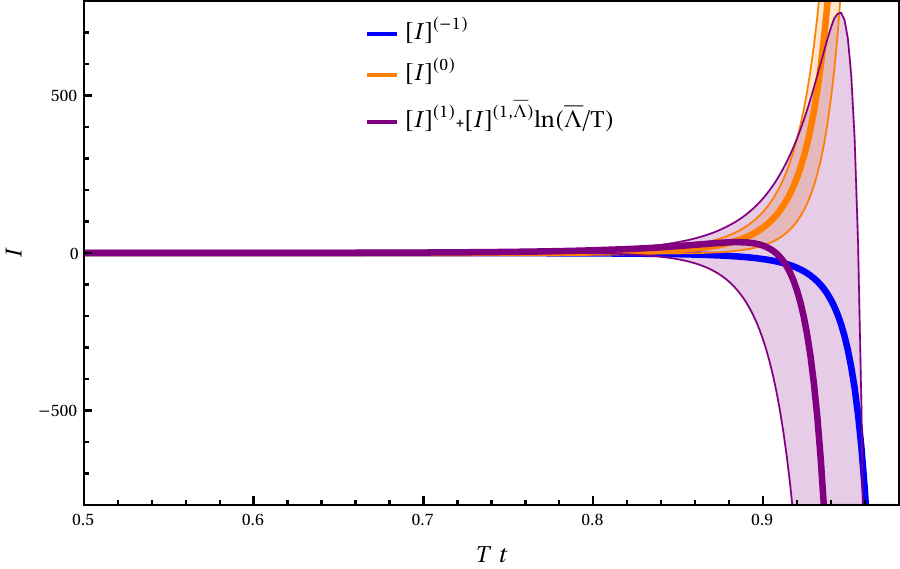}
\caption{The three components of the nonfactorisable sum-integral $I(t)$ expanded to $O(\varepsilon)$ in dimensional regularisation. 
  The bands come from variations around $\bar\Lambda_{\mathrm{PMS}}$ (for $N_f=N_c=3$).
  \label{fig:IXRfig}}
\end{figure}

The $\ln \bar \Lambda $-dependent part of the $O(\varepsilon)$ term  follows from the divergence, so we also have it in closed form, and call it $\left[I\right]^{(1,\bar \Lambda )}(t)$. 
For the complete $\left[I\right]^{(1)}(t)$ term there appears to be no simple closed-form expression, but with the $\ln \bar\Lambda$-dependence removed, the numerical integral necessary for its evaluation only depends on $Tt$. 
Parts of the integral are analytically tractable. 
For example, the antisymmetric part of the numerical integral must cancel with the explicitly subtracted $\mathcal{Z}^{1}_{011}(t)$ term. However, we have not found it particularly useful to extract such terms from it. 
While the numerical integration is somewhat delicate---the integral is, after all, only barely convergent by construction---we are able to perform it to roughly $1\%$ accuracy for $Tt<0.98$,
so that the uncertainties related to numerical precision are dwarfed by the renormalisation-scale uncertainty, even for this one contribution. 
The results for the various components of $I(t)$ are shown in figure \ref{fig:IXRfig}. 
We display only the second half of the thermal circle, as the contribution is symmetric (up to numerical accuracy).
Note that, while the band for the linear component grows large towards the edge of the thermal circle, its central numerical value is nevertheless smaller than or comparable to the other terms for $Tt \lesssim 0.92$.

\section{Explicit form of the renormalised correlators}\label{app: Correxpanded}
We write below the final expression for the renormalised correlators at NLO. 
To reproduce this form, we use eqs. \eqref{appB:EINT}, \eqref{appB:E20INT} and \eqref{appB:Ixrsubtr} and substitute them appropriately into eq.~\eqref{eq:eweresult}. 
As discussed in the main text,  we perform an expansion in $\varepsilon$ and  renormalise using eq. \eqref{gsMSbar}. At this point, all divergences are seen to cancel. 
The remaining contribution as $\varepsilon \to 0$ is:   
\begin{align}
  \langle EE\rangle_{I}= g_s^2&\left(\overline{\Lambda}\right)d_A\pi^2 \Big[\cos\left(2\pi Tt\right) + 2 \Big] \csc ^4(\pi  Tt) \nonumber\\
  &\qquad \times \bigg\{1+\frac{g_s^2}{48\pi^2}  (11N_c-2N_f)\bigg[2 \ln
   \left(\frac{\overline{\Lambda}}{4 \pi  T}\right)+1\nonumber-16\,\RePart{\,\mathrm{Li}^{(1)}_{-3}\left(e^{2 i \pi  Tt}\right)}\bigg]\bigg\} \nonumber\\
   +g_s^4& d_A N_c\Bigg\{\frac{2}{(4\pi)^4}\ln{\left[2\sin\left(\pi Tt\right)\right]}\left[\pi^2-8\ln^2\left(2\pi\right)\right]-\frac{9}{8}\csc^4\left(\pi Tt\right)\nonumber \\
   &-\csc^4\left(\pi Tt\right)\Big[\cos{\left(2\pi Tt\right)}+ 2\Big]\bigg( \frac{1}{2}\ln{\left[2\sin\left(\pi Tt \right)\right]}\nonumber\\
   &\qquad\qquad +4\gamma_E\,\RePart{\,\mathrm{Li}^{(1)}_{-3}\left(e^{2 i \pi  Tt}\right)}+8\pi^2\,\RePart{\,\mathrm{Li}^{(1)}_{1}\left(e^{2 i \pi  Tt}\right)}\bigg)\nonumber\\
   &+\csc^2\left(\pi Tt\right)\bigg(\frac{1}{12}\left[\frac{1}{2}-\gamma_E-\ln{\left(2\pi\right)}+\frac{\zeta^{\prime}(2)}{\zeta(2)}\right]+\cot^2\left(\pi Tt\right)\nonumber\\
   &\qquad\qquad +8\pi^2\,\RePart{\,\mathrm{Li}^{(1)}_{1}\left(e^{2 i \pi  Tt}\right)}-2\sin{\left(2\pi Tt\right)}\,\ImPart{\,\mathrm{Li}^{(1)}_{-2}\left(e^{2 i \pi  Tt}\right)}\nonumber\bigg)\nonumber\\
&-\frac{1}{(2\pi)^4}\,\RePart{\,\mathrm{Li}^{(2)}_{1}\left(e^{2 i \pi  Tt}\right)}+\Big(\ln\left[2\sin\left(\pi Tt\right)\right]-4\Big)\,\RePart{\,\mathrm{Li}^{(1)}_{-3}\left(e^{2 i \pi  Tt}\right)}\nonumber\\
&+\frac{1}{3}\,\RePart{\,\mathrm{Li}^{(1)}_{-1}\left(e^{2 i \pi  Tt}\right)} +2\,\ImPart{\,\mathrm{Li}^{(1)}_{0}\left(e^{2 i \pi  Tt}\right)}\nonumber\\
&-4 T \lambda (W_I)\, \ImPart{\,\mathrm{Li}^{(1)}_{-3}\left(e^{2 i \pi  Tt}\right)}-2\left[I\right]^{(1)}(t)\Bigg\}\nonumber\\
+g_{s}^4& \frac{d_AN_f}{64}\Bigg\{\csc ^4(\pi  Tt) \bigg( 8-2
   \Big[\cos \left(2 \pi  Tt \right)+3\Big] \left(1+\ln 2\right)\nonumber\\
   &\qquad\qquad +16 \sin ^3\left(\pi  Tt\right)
   \bigg[\ImPart{\,\mathrm{Li}^{(1)}_{-2}\left(e^{i \pi  Tt}\right)}-4\,
   \ImPart{\,\mathrm{Li}^{(1)}_{-2}\left(e^{2 i \pi  Tt}\right)}\bigg]\nonumber\nonumber\\
   &\qquad\qquad -8 \sin (\pi  Tt) \Big[\cos
   (2 \pi  Tt)+3\Big] \bigg[\ImPart{\,\mathrm{Li}^{(1)}_{0}\left(e^{i \pi 
   Tt}\right)}-\ImPart{\,\mathrm{Li}^{(1)}_{0}\left(e^{2 i \pi  Tt}\right)}\bigg]\bigg)\nonumber\\
   &+\frac{8}{3}\csc ^2(\pi  Tt) \bigg[-3\gamma_E +\ln (2\pi)+\frac{\zeta^{\prime}(2)}{\zeta(2)}\bigg]-\frac{32}{2}\,\RePart{\,\mathrm{Li}^{(1)}_{-1}\left(e^{2 i \pi  Tt}\right)}\Bigg\}
\end{align}
for $I\in\{ U,L\}$, where $\lambda(W_U)=t$ ($\lambda(W_{L})=-(1/T-t)$) is (minus) the length of the Wilson line for the upper (lower) correlator, $\gamma_E$ is the Euler constant and $\left[I\right]^{(1)}(t)$ is the numerically evaluated integral following the discussion in Appendix \ref{app:ixr}. 
We have used $ \mathrm{Li}^{(n)}_{s}\left(e^{ i \pi  x}\right)+\mathrm{Li}^{(n)}_{s}\left(e^{ -i \pi  x}\right)= 2\,\RePart{\,\mathrm{Li}^{(n)}_{s}\left(e^{ i \pi  x}\right)}$ and $ \mathrm{Li}^{(n)}_{s}\left(e^{ i \pi  x}\right)-\mathrm{Li}^{(n)}_{s}\left(e^{ -i \pi  x}\right)= 2i\,\ImPart{\,\mathrm{Li}^{(n)}_{s}\left(e^{ i \pi  x}\right)}$ with $x$ either $Tt$ or $2Tt$. 
We note that at least some derivatives of polylogarithms can be written in terms of generalised zeta functions,  
e.g. for the coefficient of the term responsible for the asymmetry it holds that 
$\ImPart{\,\mathrm{Li}^{(1)}_{-3}\left(e^{2 i \pi  Tt}\right)=-3\left[\zeta\left(4,Tt\right)-\zeta\left(4,1-Tt\right)\right]/\left(16\pi^3\right)}$, 
see eq. \eqref{eq:zresult}. 
\bibliographystyle{jhep}
\bibliography{references}

\end{document}